\documentclass[preprint2]{aastex}

\usepackage{natbib}

\slugcomment{To appear in the Astrophysical Journal}

\shorttitle{VLA Observations of BLAST Cores}
\shortauthors{Olmi et al.}

\begin{document}


\title{ High Angular Resolution Observations of Four Candidate BLAST High-Mass Starless Cores.
}

\author{Luca Olmi,\altaffilmark{1,2,\dag} 
        Esteban D. Araya,\altaffilmark{3}     
        Edward L. Chapin,\altaffilmark{4}
        Andrew Gibb,\altaffilmark{4}
	Peter Hofner,\altaffilmark{5,6,7}
        Peter G. Martin,\altaffilmark{8,9}
	Carlos M. Poventud\altaffilmark{1}}

\altaffiltext{1}{University of Puerto Rico, Rio Piedras Campus, Physics Dept., Box 23343, UPR station, San Juan, Puerto Rico}

\altaffiltext{2}{INAF, Osservatorio Astrofisico di Arcetri, Largo
E. Fermi 5, I-50125, Firenze, Italy.}

\altaffiltext{3}{Physics Department, Western Illinois University, 1 University Circle, Macomb, IL 61455, USA}

\altaffiltext{4}{Department of Physics \& Astronomy, University of
British Columbia, 6224 Agricultural Road, Vancouver, BC V6T~1Z1,
Canada}

\altaffiltext{5}{Physics Department, New Mexico Institute of Mining and Technology, 
801 Leroy Place, Socorro, NM 87801, USA}

\altaffiltext{6}{NRAO, P.O. Box O, 1003 Lopezville Road, Socorro, NM 87801-0387, USA    }

\altaffiltext{7}{Max-Planck-Institut f\"ur Radioastronomie, Auf dem H\"ugel 69, 53121 Bonn, Germany    }

\altaffiltext{8}{Canadian Institute for Theoretical Astrophysics, University of Toronto, 60 St. George Street, 
Toronto, ON M5S~3H8, Canada}

\altaffiltext{9}{Department of Astronomy \& Astrophysics, University of Toronto, 50 St. George Street, 
Toronto, ON  M5S~3H4, Canada}

\altaffiltext{\dag}{\url{olmi.luca@gmail.com, olmi@arcetri.astro.it}}

\begin{abstract}
We discuss high-angular resolution observations of ammonia toward four candidate high-mass starless
cores (HMSCs). The cores were identified by the Balloon-borne Large Aperture Submillimeter Telescope
(BLAST) during its 2005 survey of the Vulpecula region where   
60 compact sources were detected simultaneously at 250, 350, and 500\,\micron.
Four of these cores, with no {\it IRAS}-PSC or {\it MSX} counterparts, 
were mapped with the NRAO Very Large Array (VLA) and observed with the Effelsberg 100\,m telescope
in the NH$_3$(1,1) and (2,2) spectral lines. 
Our observations indicate that the four cores are cold ($T_{\rm k} < 16\,$K) and show
a filamentary and/or clumpy structure. They also show a significant velocity substructure 
within $\sim 1\,$km\,s$^{-1}$.  The four BLAST cores appear to be colder
and more quiescent than other previously observed HMSC candidates, suggesting an
earlier stage of evolution.  

\end{abstract}


%
\begin{deluxetable}{lccccc}
\tablewidth{0pt}
\small
\tablecaption{VLA Observing Parameters
 \label{tab:vla}}
\tablehead{
\colhead{BLAST ID} &
\colhead{Source Name} &
\colhead{RA[J2000]} &
\colhead{DEC[J2000]} &
\colhead{Synthesized beam} &
\colhead{Sensitivity} \\
\colhead{} &
\colhead{} &
\colhead{} &
\colhead{} &
\colhead{arcsec} &
\colhead{mJy\,beam$^{-1}$}
}
\startdata
V10  & BLAST J194106+235513  & 19:41:06.5  & +23:55:13.5  & $3.3 \times 2.9$  & 4.7 \\   
V11  & BLAST J194136+232325  & 19:41:36.3  & +23:23:24.9  & $3.3 \times 2.8$  & 4.3 \\   
V27  & BLAST J194306+230125  & 19:43:06.5  & +23:01:25.5  & $3.4 \times 2.7$  & 4.4 \\   
V33  & BLAST J194319+232639  & 19:43:19.0  & +23:26:39.6  & $3.4 \times 3.0$  & 5.9 \\   
\enddata
\tablecomments{
The coordinates represent the BLAST core peak positions and VLA phase tracking center.
The synthesized beam and sensitivity refer to the NH$_3$(1,1) line.
}
\end{deluxetable}

\section{INTRODUCTION }

The importance of massive ($M \ga 8\, {\rm M}_\odot$) stars in shaping the Galactic structure
and evolution is well known \citep{zinnecker2007}.
In recent years, a major observational effort has been made to identify the earliest stages
of their evolution, which are not very well constrained, yet.
Systematic studies by various groups have uncovered several high-mass
proto-stellar objects (HMPOs), i.e.,  dense gravitationally bound cores
in a pre-ultracompact (UC) H{\sc II} region phase with typical temperatures
$T\sim 15-90\,$K (\citealp{mol96}, \citealp{molinari2002}, \citealp{Sridharan02},
\citealp{beu02}).

However, these surveys were carried out on {\it IRAS}-selected objects
and thus there are very few observations of
the so-called high-mass starless (or pre-protostellar) core (HMSC)
stage, which is supposed to precede the HMPO phase,
and their physical and kinematical properties have been barely studied.
The survey by \citet{Sridharan05} was also based on previous observations of {\it IRAS}-selected
candidate HMPOs at 1.2\,mm, thus quite far from the emission peak of the spectral energy
distribution (SED) of the coldest ($T\la 15\,$K) cores.
The survey of infrared dark clouds (IRDCs) by \citet{pillai2006} showed that
IRDCs are cold ($T < 20\,$K) massive ($M>100\,$M$_\odot$) and have linewidths
$\simeq 1-3\,$km\,s$^{-1}$. However, these authors probed regions
with typical sizes $\ga 1\,$pc, thus likely to represent pre-protoclusters.
\citet{motte2007} in their  unbiased survey of Cygnus~X  found 17 cores qualifying
as good candidates for hosting massive IR-quiet protostars, driving outflows
traced by SiO emission, but failed to discover the high-mass analogs of pre-stellar dense cores.

Detailed studies of individual objects (e.g., G28.34+0.06, \citealp{wang2008},  \citealp{zhang2009};
IRAS 05345+3157,  \citealp{fontani2009}; Cygnus~X, \citealp{bontemps2009}) 
are now contributing to set further constraints on the fragmentation inside 
massive dense clumps, leading to individual collapsing protostars. These works have 
tentatively identified cores in very early evolutionary stages, which make them  good
targets to study the initial fragmentation phases in molecular clumps.  
In this paper we describe four massive cores that appear to be even colder and more quiescent 
than similar objects observed in previous works, and may thus further constitute excellent targets
for follow-up studies on the initial fragmentation phases.

The {\it Balloon-borne Large Aperture Submillimeter Telescope} (BLAST) has
recently identified a new and unique sample of massive, cold
dust clumps, with characteristics sizes $\la 0.4\,$pc.
BLAST is a 2-m stratospheric balloon telescope that observes simultaneously at
250, 350, and 500\,\micron\ using bolometric imaging arrays
\citep{pascale2008}.  BLAST, until the first Galactic results from {\it Herschel}
become available, is unique in its ability to
detect and characterize cold dust emission from both starless and protostellar
sources, constraining the temperatures of objects with $T \la 25$\,K
using its three-band photometry near the peak of the cold core SED.
During the first BLAST science flight (BLAST05), BLAST
conducted the first sensitive large-scale Galactic Plane surveys at
these wavelengths.

One of the regions observed by BLAST05 covered
4\,deg$^2$ near the open cluster NGC~6823 in the constellation
Vulpecula ($\ell=59^\circ$), at a distance of about 2.3\,kpc (see discussion
in \citealp{chapin2008}). In this region,
60 compact sources ($<60''$ diameter) were detected simultaneously in all
three bands.  Their SEDs were
constrained through BLAST, {\it IRAS}, {\it Spitzer} MIPS and {\it
MSX} photometry, with inferred dust temperatures spanning
$\sim 12$--40\,K assuming a dust emissivity index $\beta=1.5$,
and total masses $\sim 15$--700\,$M_\odot$ \citep{chapin2008}.
At least 30\% of these cores are new, with no {\it IRAS}-PSC or {\it MSX} associations.
Even then, most of those with {\it IRAS} identifications have not been studied in detail.
The sources detected in the BLAST bands, in particular those
without {\it IRAS} counterparts, indicate the presence of significant
quantities of cool dust.  Thus, the BLAST observations resulted in an unique and
uniform sample, which may contain a number of {\it bona-fide} HMSCs.

Among the 60 compact BLAST sources, we selected the four coldest, massive and IR-quiet cores 
(see Section~\ref{sec:sample}) for observations with the NRAO\footnote{
The National Radio Astronomy Observatory is a facility of the
National Science Foundation operated under cooperative agreement
by Associated Universities, Inc.} Very Large Array (VLA)
in the NH$_3$(1,1) and (2,2) spectral lines.
The structure of the paper is thus the following: we describe the observations and discuss the
main results in Section~\ref{sec:obs}. Our analysis is presented in  Section~\ref{sec:analysis} and finally
our conclusions are summarized in Section~\ref{sec:concl}.

%
 \begin{figure}
 \centering
 \hspace{-0.3cm}
 \includegraphics[width=7.0cm]{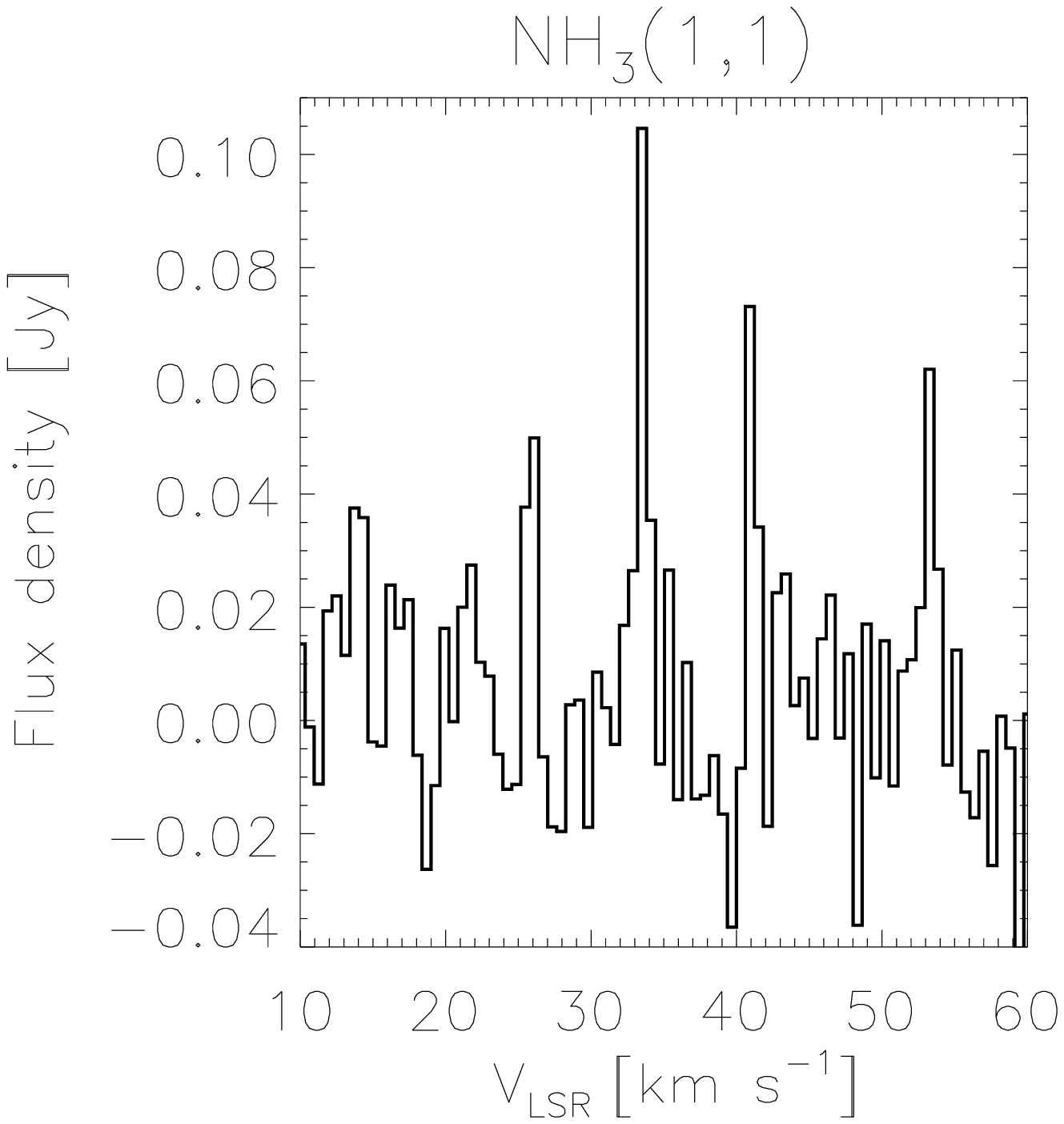}
 \includegraphics[width=4.6cm,angle=270]{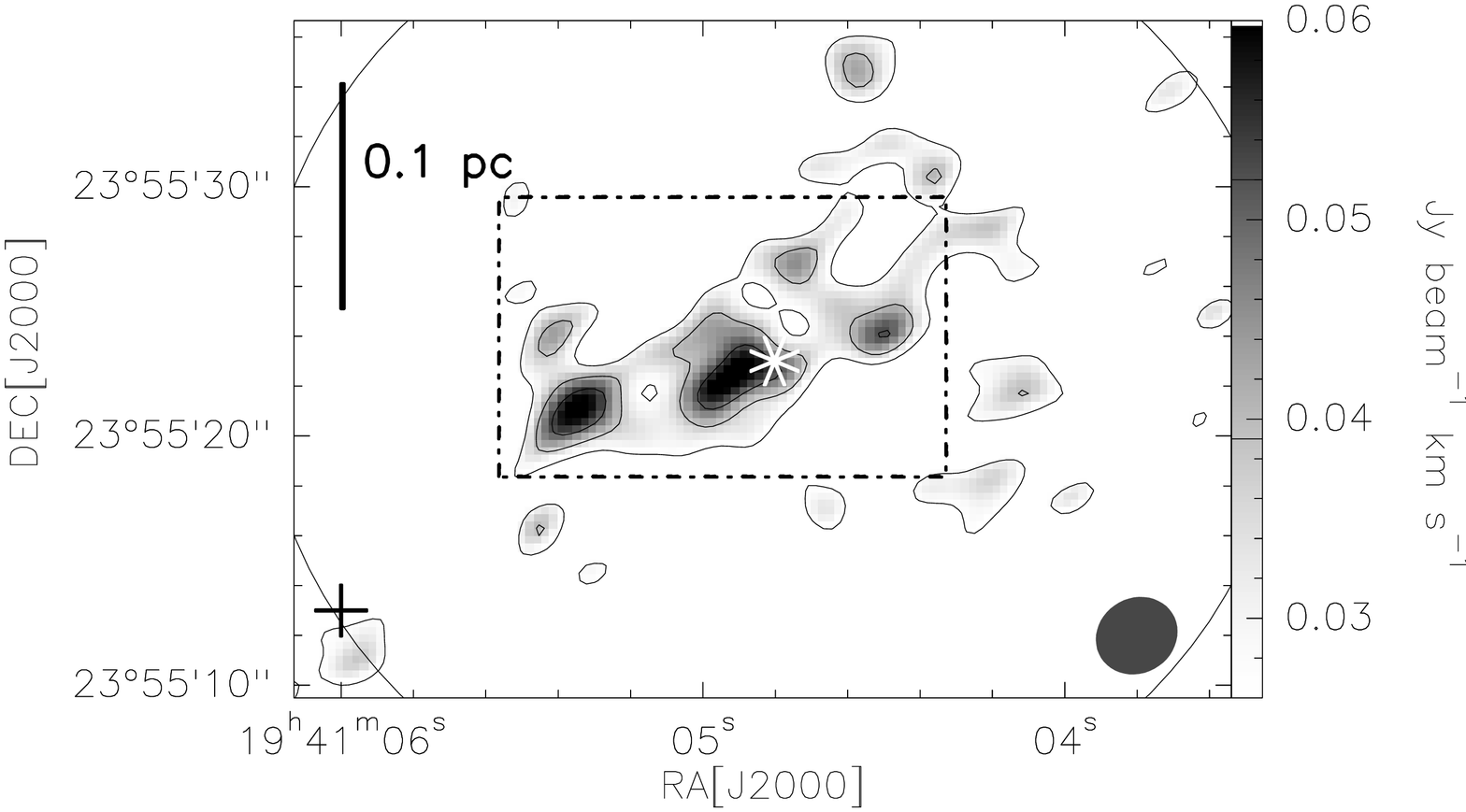}
 \vspace{0.5cm}
 \caption{
{\bf V10}. {\it Top.} NH$_3$(1,1) VLA integrated spectrum of V10.
{\it Bottom.} VLA map of the integrated intensity of NH$_3$(1,1)
over main and satellite components. The first contour level is at +2-$\sigma$
and the interval between adjacent levels is 1-$\sigma$ (RMS~$=13\,$mJy\,beam$^{-1}$\,km\,s$^{-1}$).
The values corresponding to
the contour levels are also indicated by the horizontal lines drawn in the
grey scale wedge to the right.  The ellipse in the bottom right hand corner represents the half
power width of the synthesized beam, while the cross indicates the nominal position
of the BLAST core (whose uncertainty is $\sim 1\,$arcmin) 
and the phase tracking center. The pointing position of the 100-m
observations is shown by the asterisk and the large solid circle (only partially visible)
represents the half power width of the 100-m beam.
The dot-dashed box represents the area where the line emission has been integrated to generate
the spectrum shown in the top panel.
The linear scale in the source is shown by the vertical bar in the left hand side.
}
\label{fig:v10}
\end{figure}

%
%
%
%
 \begin{figure*}
 \centering
\hspace{-2.2cm}
\includegraphics[width=7.3cm,angle=0]{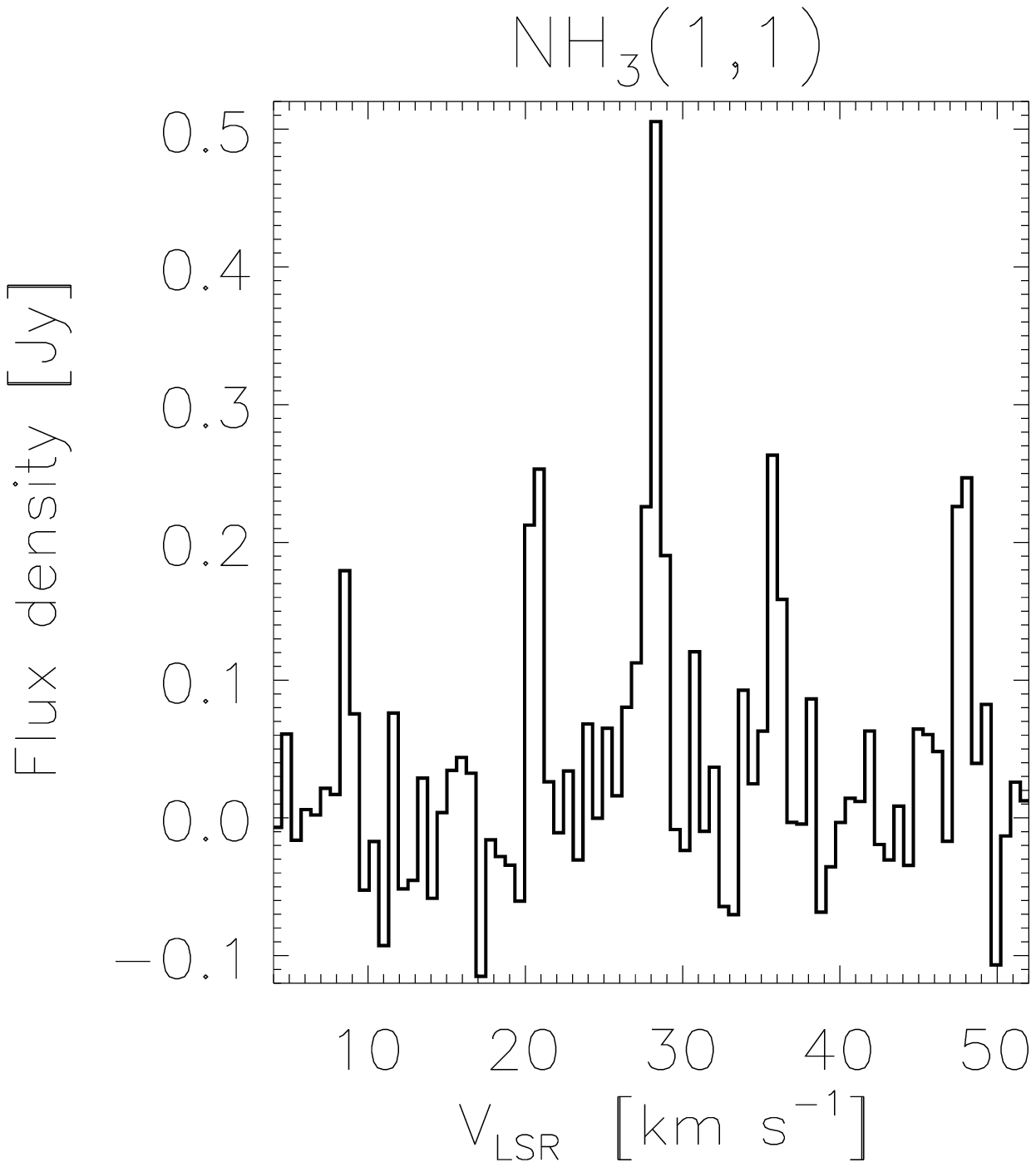}
\hspace{0.4cm}
\includegraphics[width=7.3cm,angle=0]{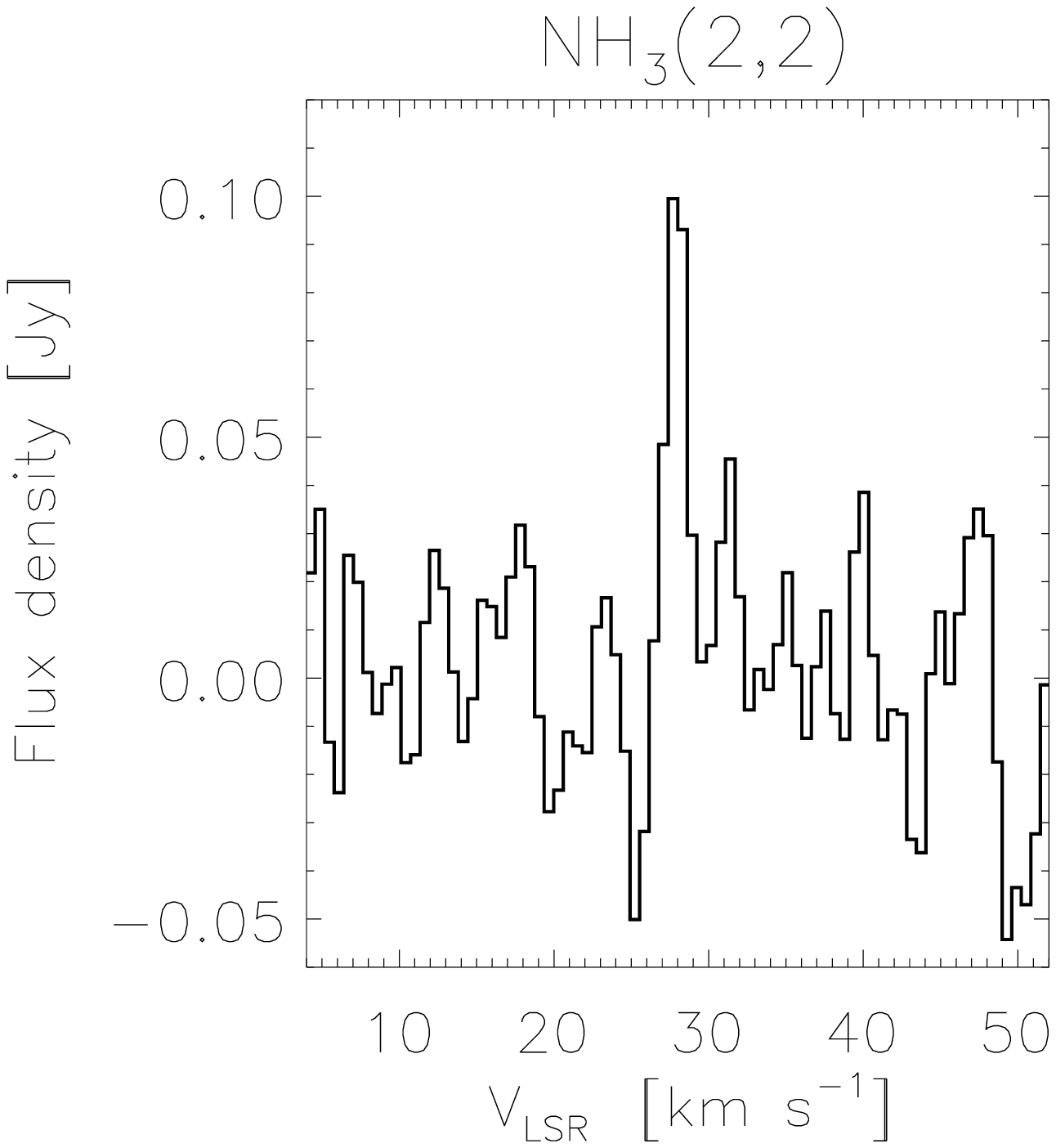}
%
\includegraphics[width=5.3cm,angle=270]{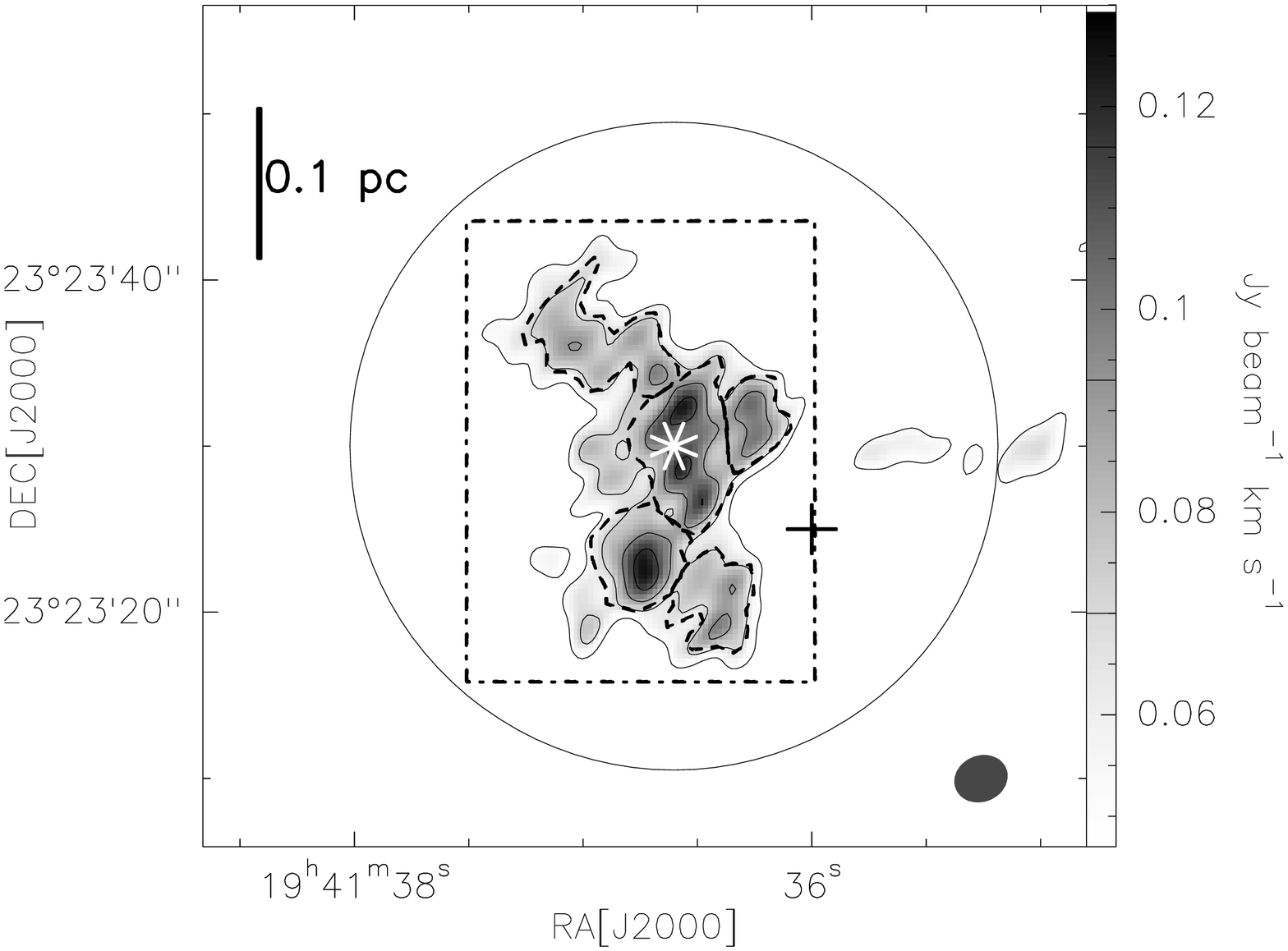}
\hspace{0.7cm}
\includegraphics[width=5.7cm,angle=270]{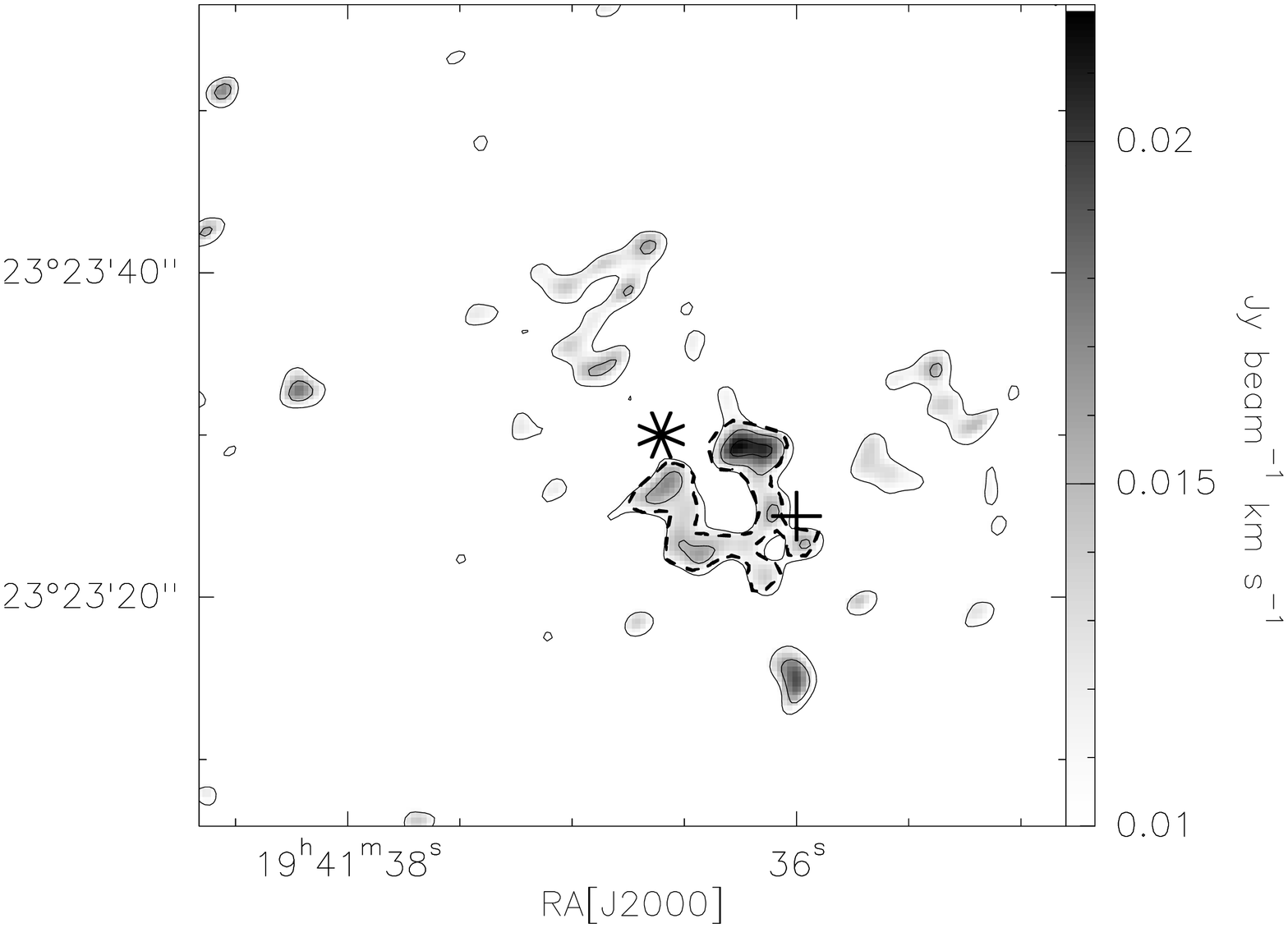}
 \caption{
{\bf V11}. {\it Top.} NH$_3$(1,1) (left) and (2,2) (right) VLA integrated spectra of V11.
{\it Bottom.} VLA maps of the NH$_3$(1,1) (left) and (2,2) (right) lines, integrated
over the main and satellite (in the case of the (1,1) line) components.
The boldface dashed contours in the (1,1) map indicate the specific sub-regions used to calculate
the total mass of the source (see Section~\ref{sec:mass}). The boldface dashed contour in the
(2,2) map indicates the region where emission has been averaged to produce the spectrum
shown in Figure~\ref{fig:22mask}.  Other contours and features are as in Figure~\ref{fig:v10}
(with RMS~$=23\,$mJy\,beam$^{-1}$\,km\,s$^{-1}$ and $5\,$mJy\,beam$^{-1}$\,km\,s$^{-1}$
in the (1,1) and (2,2) maps, respectively).
}
\label{fig:v11}
\end{figure*}

%
\begin{figure*}
 \centering
\hspace{-1.1cm}
\includegraphics[width=7.3cm,angle=0]{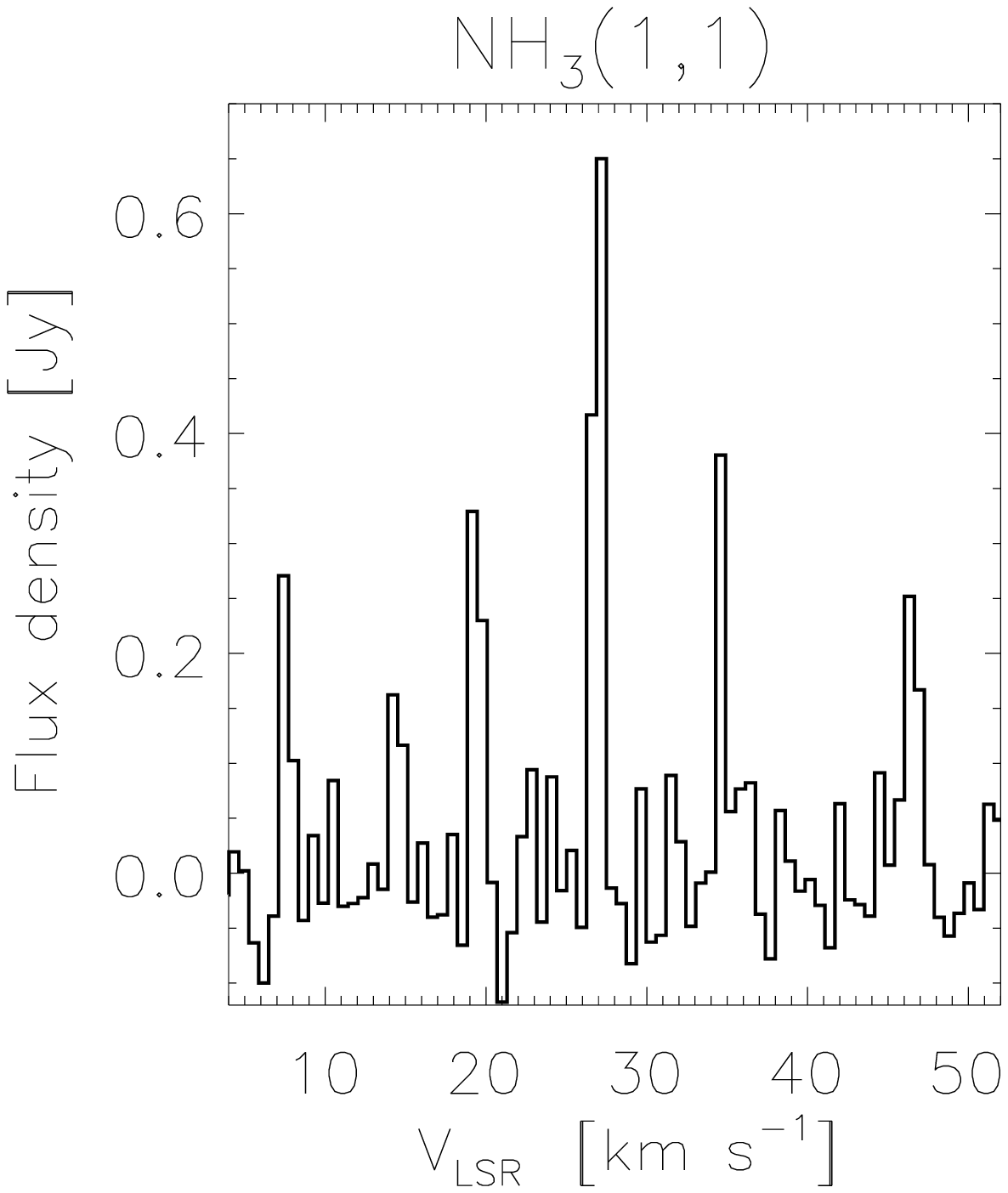}
\hspace{0.7cm}
\includegraphics[width=7.3cm,angle=0]{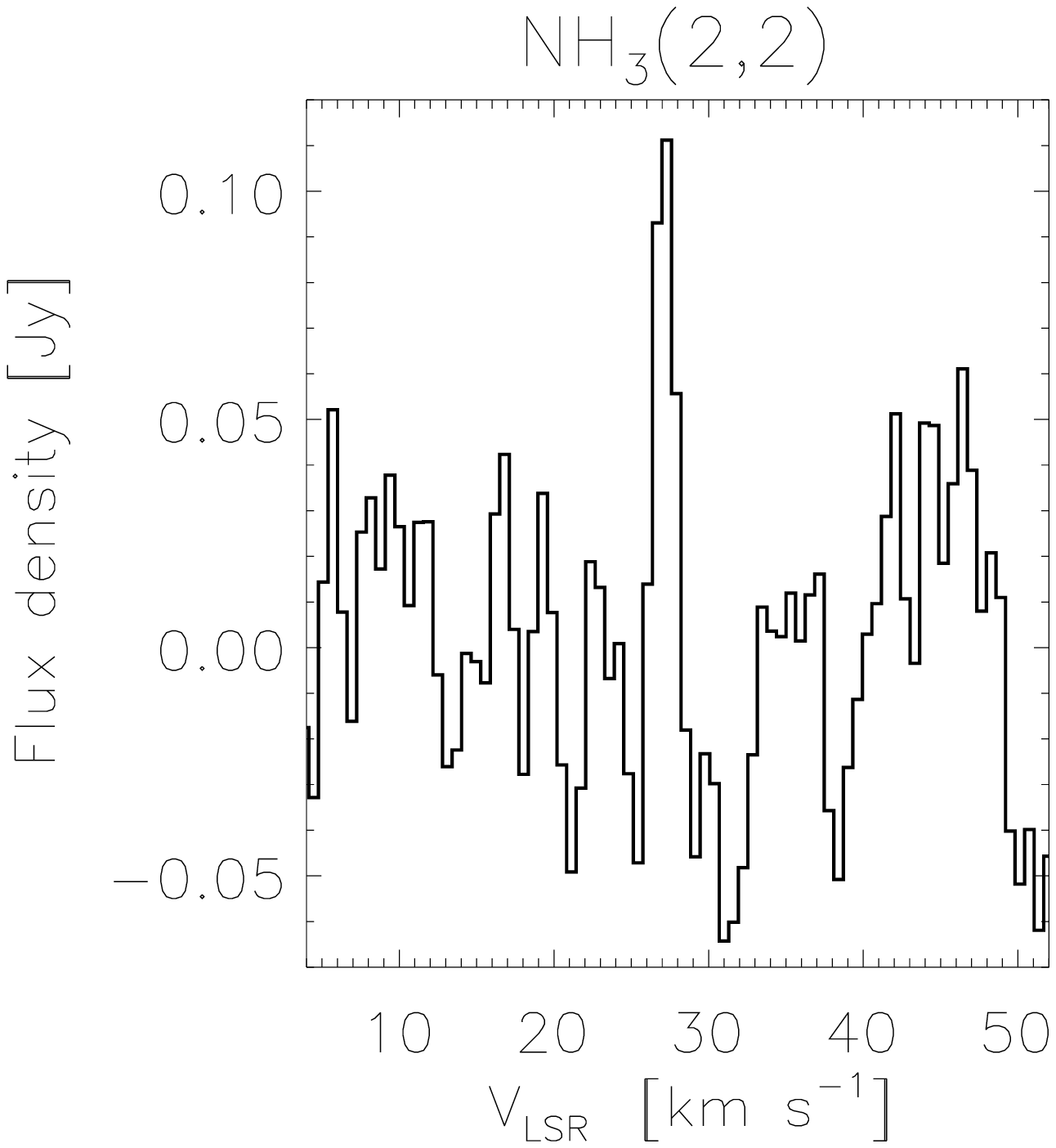}
\includegraphics[width=5.7cm,angle=270]{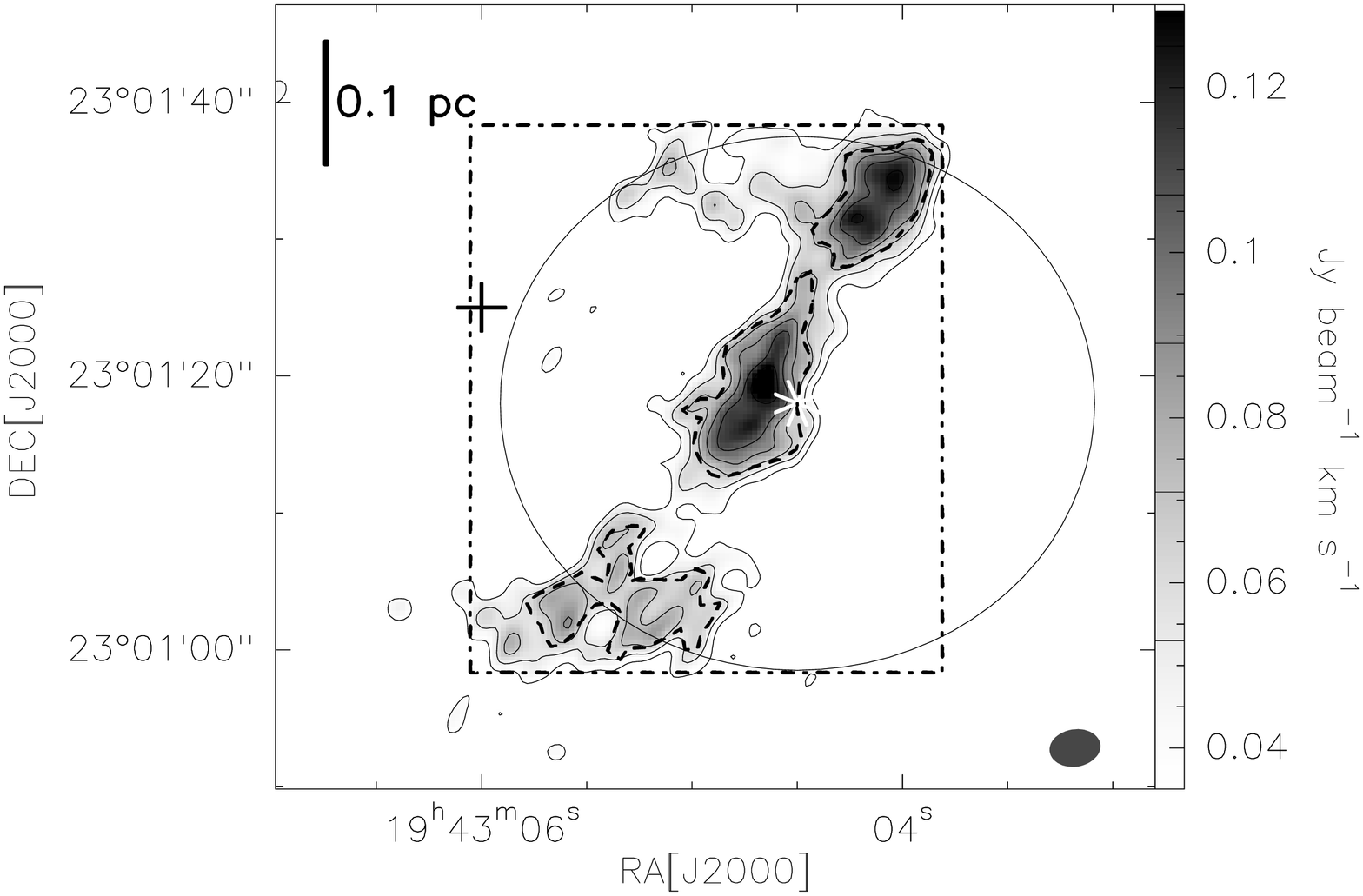}
\includegraphics[width=5.7cm,angle=270]{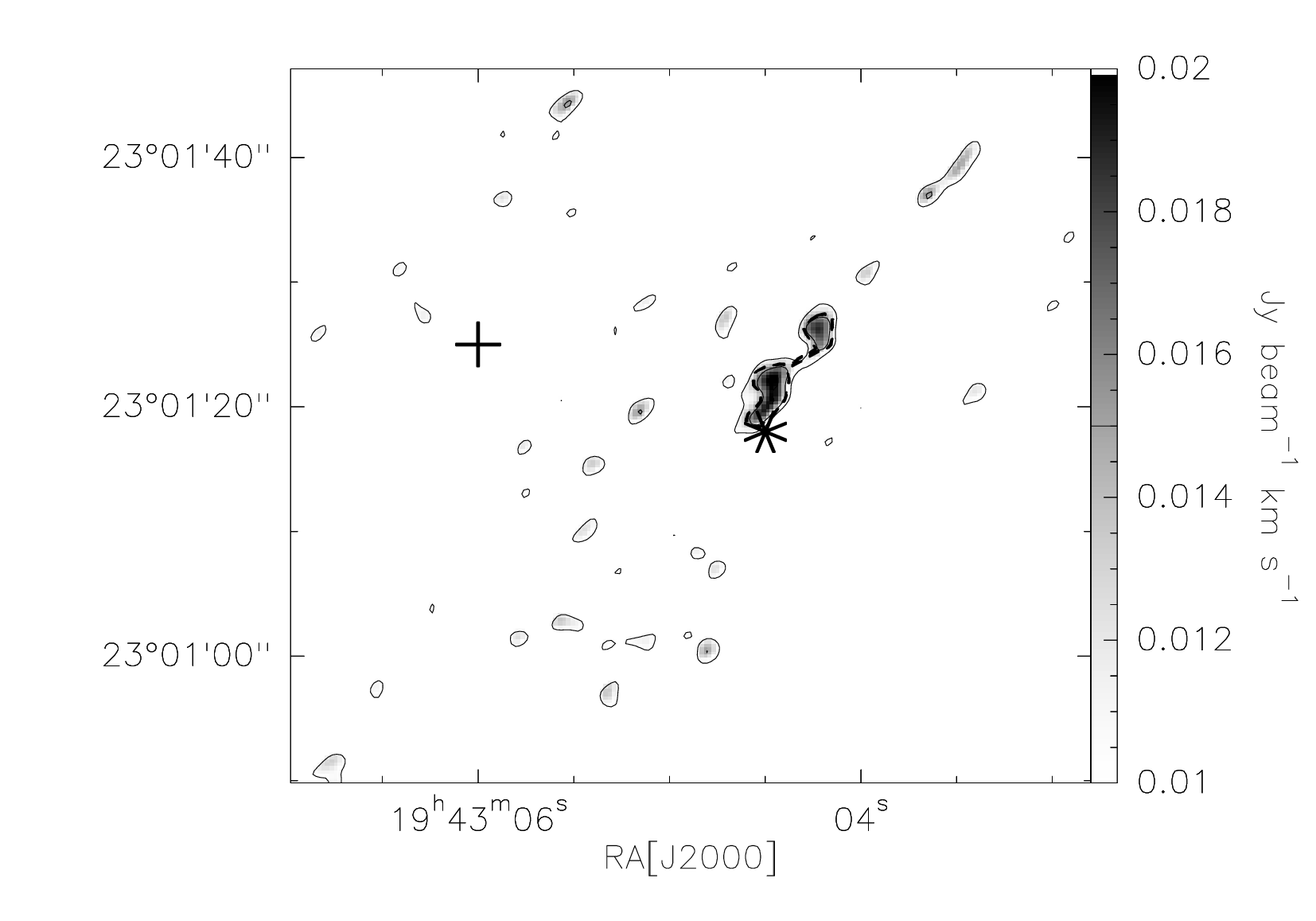}
 \caption{
{\bf V27}. Same as Figure~\ref{fig:v11} for V27 (with RMS~$=18\,$mJy\,beam$^{-1}$\,km\,s$^{-1}$
and $5\,$mJy\,beam$^{-1}$\,km\,s$^{-1}$ in the (1,1) and
(2,2) maps, respectively). }
\label{fig:v27}
\end{figure*}

%
 \begin{figure*}
 \centering
\epsscale{2.1}
\hspace{-1.5cm}
\plottwo{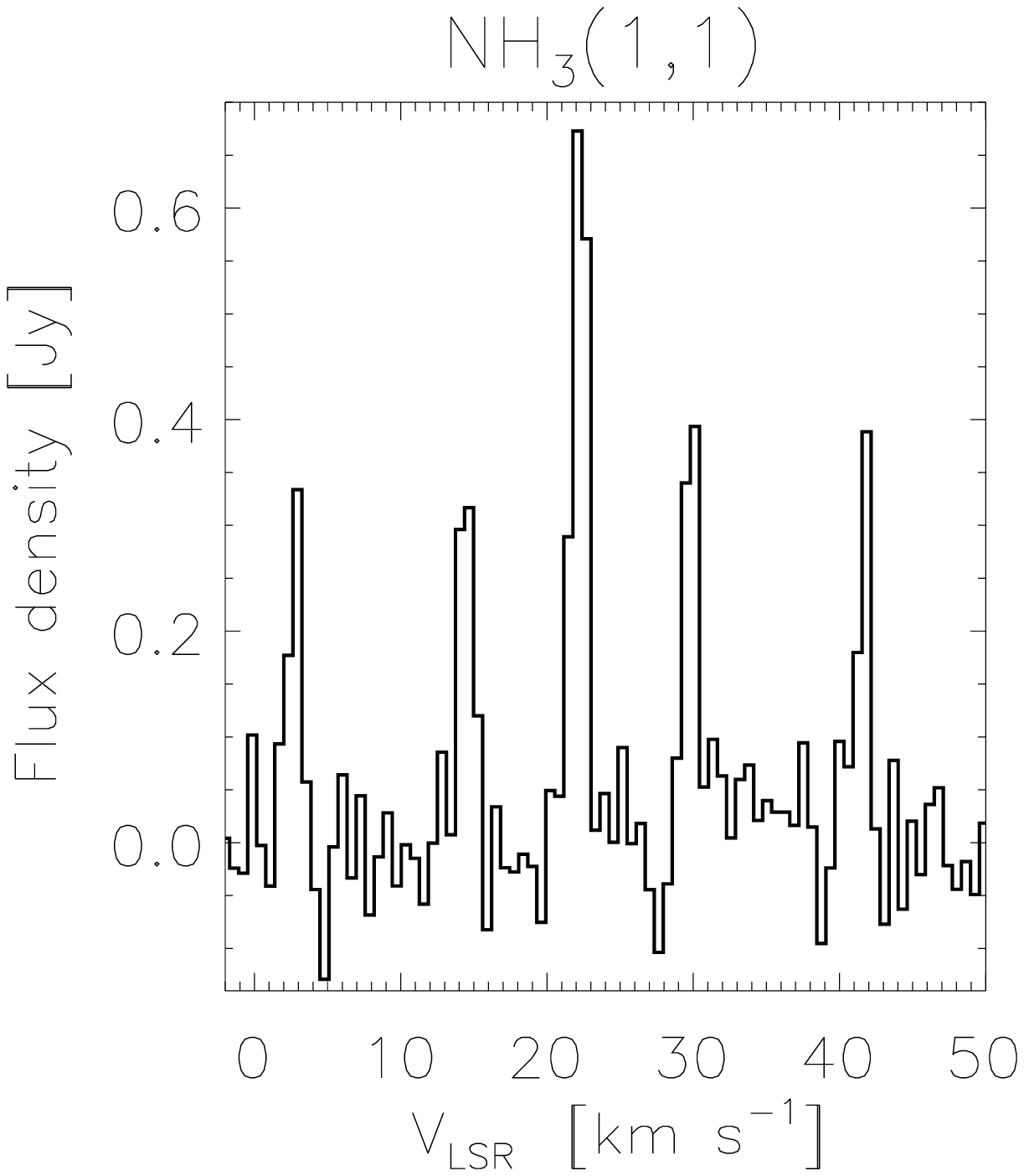}{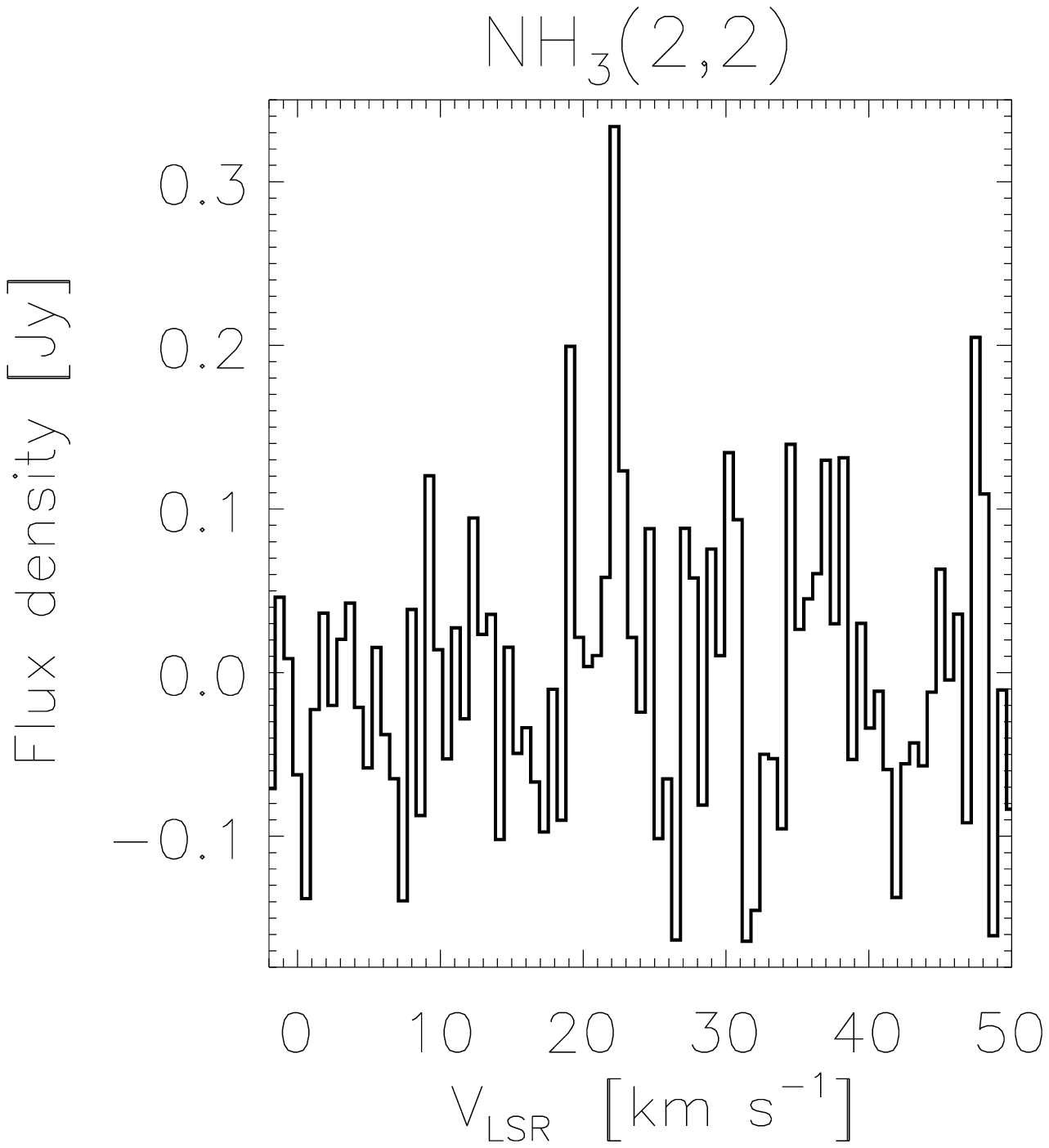}
\includegraphics[width=5cm,angle=270]{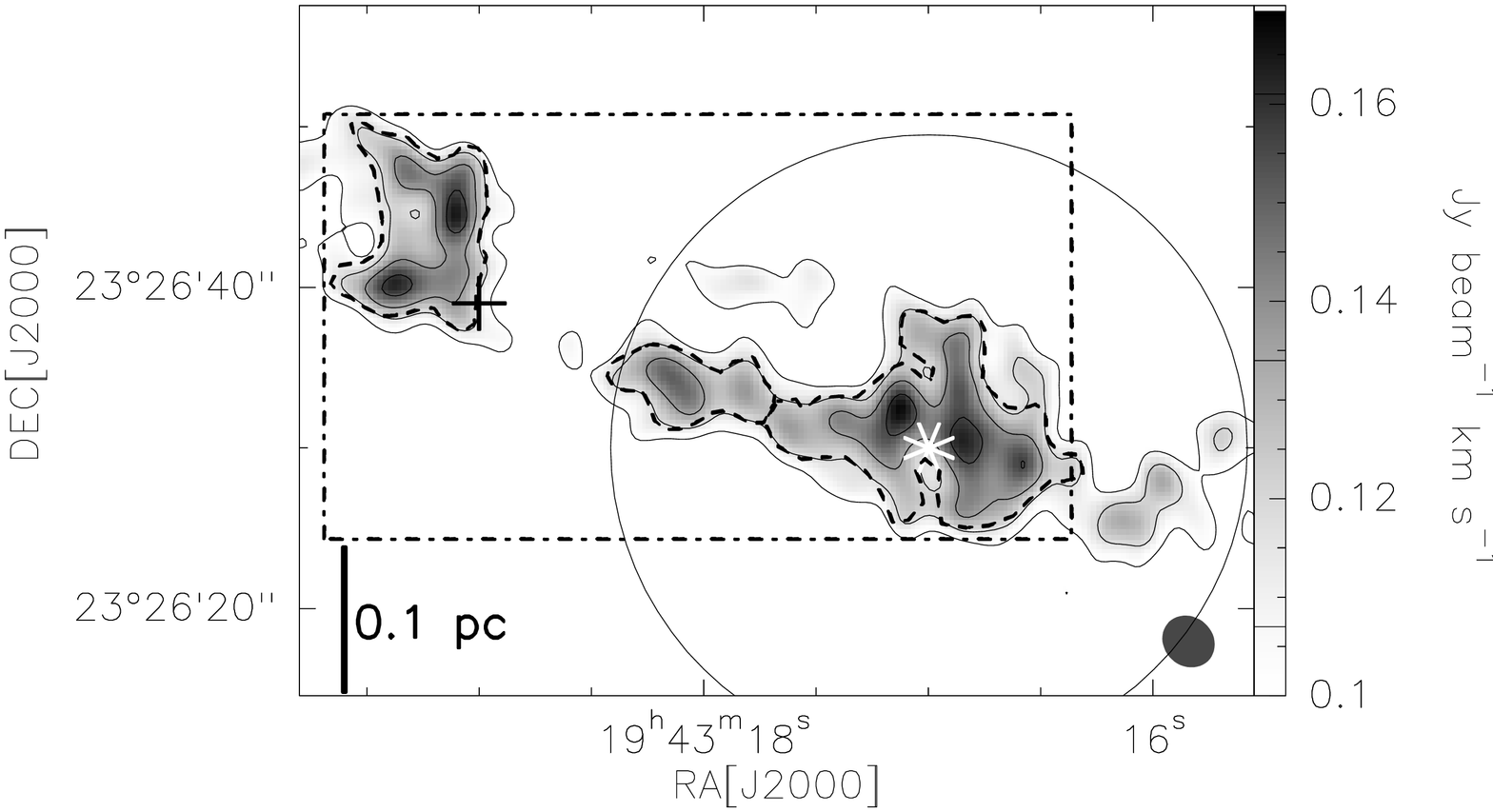}
\vspace{0.5cm}
 \caption{
{\bf V33}. Same as Figure~\ref{fig:v11} for V33 (with RMS~$=27\,$mJy\,beam$^{-1}$\,km\,s$^{-1}$ 
in the (1,1) map),
except for the first contour level that here is at 3-$\sigma$.  In the bottom
panel only the map of the integrated (1,1) emission is shown (see text).
}
\label{fig:v33}
\end{figure*}

\section{OBSERVATIONS AND DATA REDUCTION}
\label{sec:obs}

\subsection{The Sample}
\label{sec:sample}

Among the starless cores candidates in Vulpecula, i.e. with no
{\it IRAS}-PSC or {\it MSX} counterparts, we selected the four coldest ($T_{\rm dust} < 16\,$K) BLAST sources,
V10, V11, V27 and V33 (see Table~\ref{tab:vla}), to be observed with the VLA.
An analysis of {\it Spitzer} IRAC and MIPS images reveals that only source V11 might have an associated
protostar, as we discuss later (Section~\ref{sec:evo}).

These cores are also massive (with envelope masses $\sim 90$--200\,$M_\odot$) and both their
absolute luminosities ($L_{\rm FIR} \sim 50\,L_\odot$) and luminosity-to-mass ratios
($L_{\rm FIR}/M \la 1\,L_\odot M_\odot^{-1}$, see Figure~18 of \citealp{chapin2008})
suggest an early phase of evolution. Qualitatively, this can also be seen by positioning
the four sources in the $L-M$ diagram  of \citet{molinari2008} (see their Figure~9), where they would
fall in the region representing the early accretion phase. On the same diagram, it can be seen
that these BLAST cores are clearly separated from the low-mass regime.

None of the selected sources in Vulpecula is listed in the catalog of Extended Green Objects (EGOs)
of \citet{cyganowski2008}, a sample of massive young stellar objects outflow candidates
extracted from the GLIMPSE survey \citep{ben03}.  Among the
Vulpecula sources that fall in the GLIMPSE survey area (which include V10, V11, V27 and V33) only
V09 qualifies as an EGO. This is yet another indication of the early evolutionary phase of
the BLAST cores in our sample.

These four Vulpecula cores were observed with the VLA
in the NH$_3$(1,1) and (2,2) spectral lines, which are low-excitation molecular lines,
thus appropriate to study the cold cores in the BLAST catalog. In addition, the NH$_3$ molecule
is  not expected to be much depleted in pre-stellar cores (e.g., \citealp{aikawa2005}),
although its exact abundance may depend on the core density (\citealp{tafalla2002}, \citealp{flower2006}).

\subsection{VLA Observations}
\label{sec:vla}

The VLA ammonia observations were carried out in June 2008, and the
array was used in its most compact configuration (D), with baselines from 35\,m to 1\,km.
The NH$_3$(1,1) and (2,2) inversion lines at 23.694496
and 23.722633 GHz, respectively \citep{ho1983}, were simultaneously
observed in the 1IF mode, with a bandwidth of $6.25\,$MHz and 128 channels, corresponding to
a velocity coverage of about $80\,$km\,s$^{-1}$ and a velocity resolution of
about $0.6\,$km\,s$^{-1}$.
The mapped area in each source was $\simeq 2.5\arcmin \times 2.5\arcmin$, centered around
the nominal positions of the BLAST cores shown with crosses in Figures~\ref{fig:v10} to ~\ref{fig:v33} and
Figures~\ref{fig:v11ch} and \ref{fig:v27ch}, and with a synthesized beam FWHM $\simeq 3\,$arcsec
(see Table~\ref{tab:vla}).
The total time on-source varied from $\simeq 30$ to $\simeq 40\,$min per source and per line.
The flux density scale was established by observing 3C48 
and the phase calibration was ensured by frequent observations of the point source
J19259+21064, that had a measured flux density at the time of the
observations of 2.1\,Jy.
The data were edited and calibrated using the Astronomical
Image Processing System (AIPS) following standard procedures.
Imaging and deconvolution was performed using the
IMAGR task and naturally weighting the visibilities.
The resulting observing parameters are listed in Table~\ref{tab:vla}.

%
 \begin{figure}
 \centering
 \includegraphics[width=7cm]{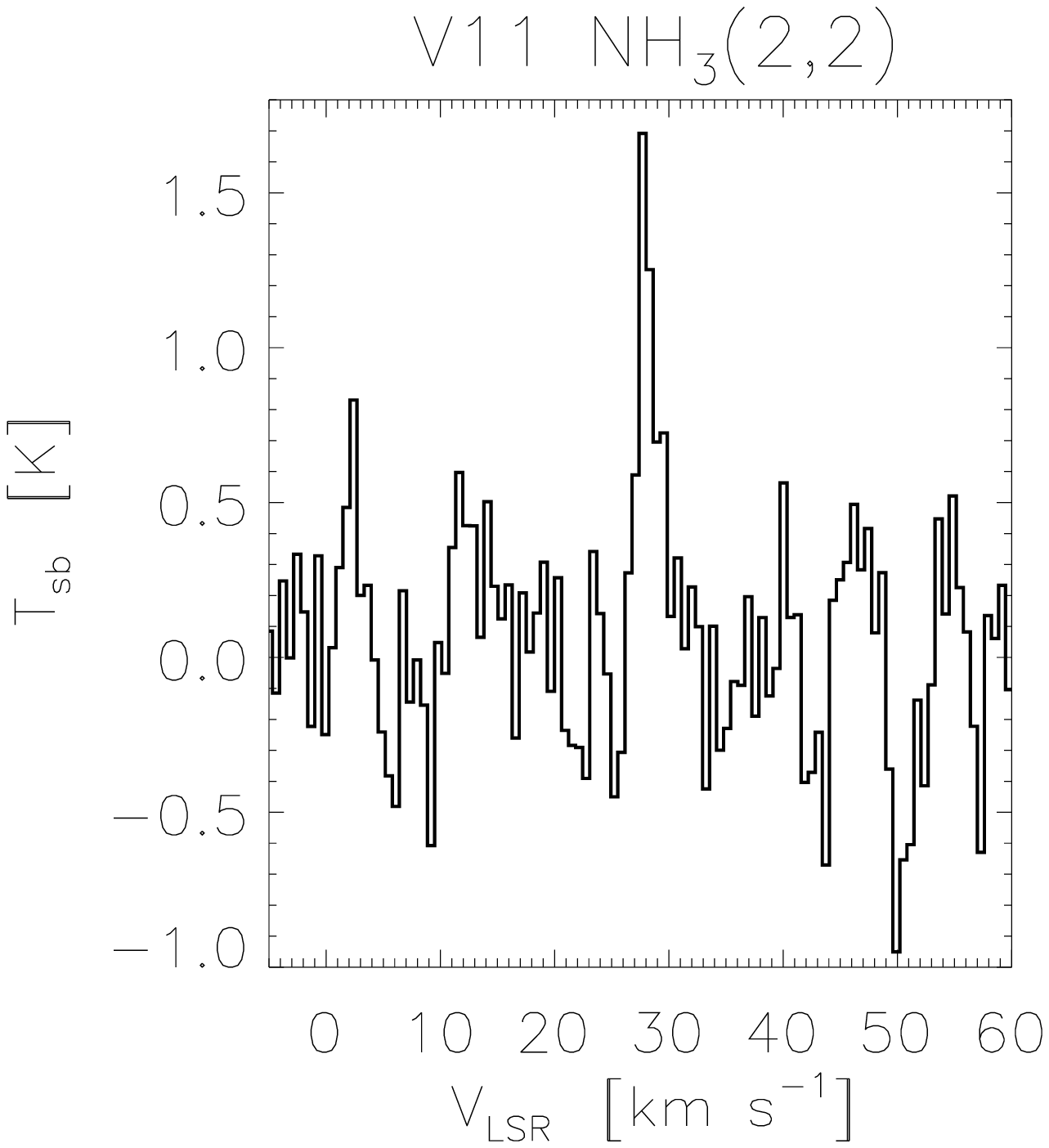}
 \includegraphics[width=7cm]{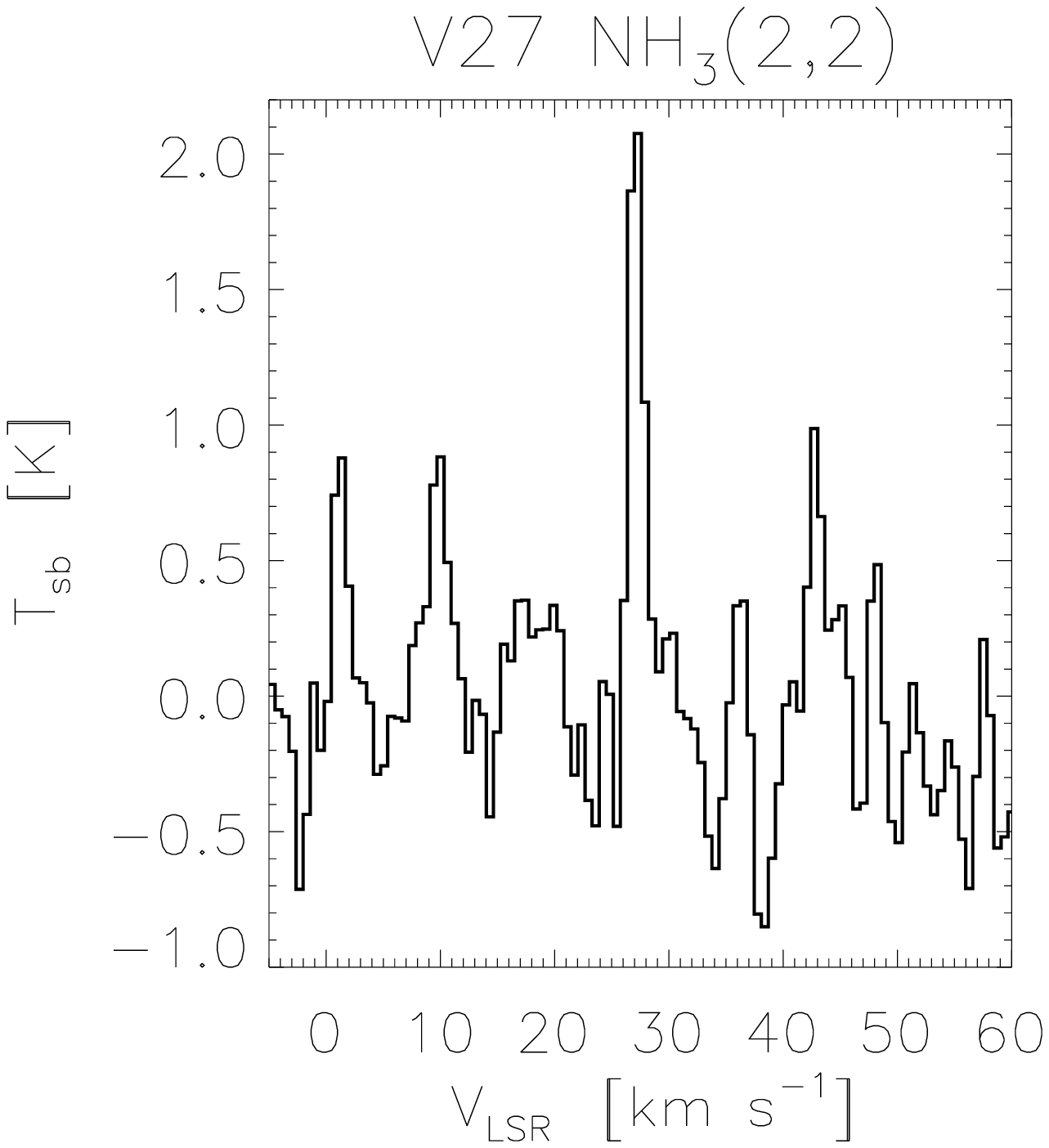}
 \caption{
NH$_3$(2,2) VLA spectra toward V11 (top panel) and V27 (bottom panel), obtained by
integrating the emission within the boldface dashed contours shown in the (2,2) maps
of Figures~\ref{fig:v11} and \ref{fig:v27}.
}
\label{fig:22mask}
\end{figure}

%
 \begin{figure*}
 \centering
 \includegraphics[width=7.5cm,angle=270]{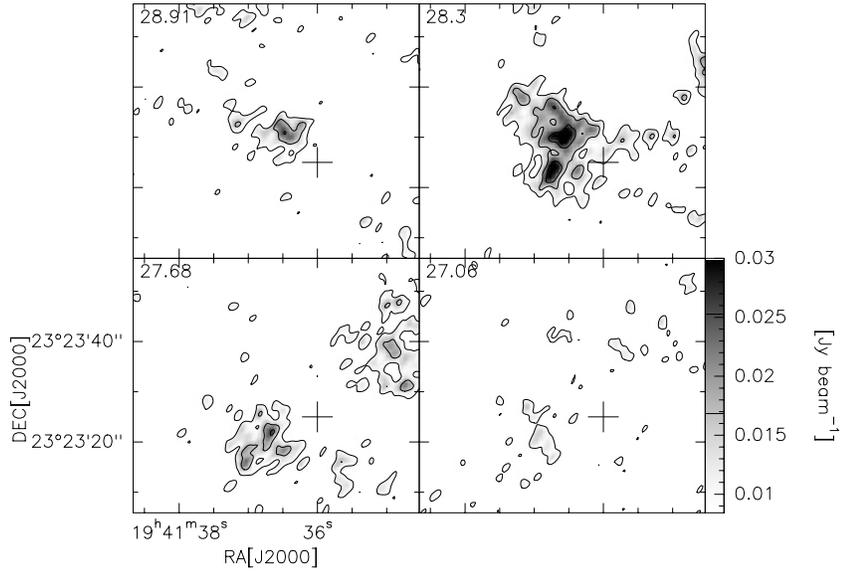}
\vspace{0.2cm}
 \caption{
{\bf V11}. VLA map of NH$_3$(1,1) main line toward V11, in four adjacent velocity channels
(shown in km\,s$^{-1}$ in the top left corner of each panel).
The lowest contour level is 2-$\sigma$. The insets  show
the synthesized beam at half power and the cross indicates the nominal position
of the BLAST core and the phase tracking center.
}
\label{fig:v11ch}
\end{figure*}

%
 \begin{figure*}
 \centering
 \includegraphics[width=8.5cm,angle=270]{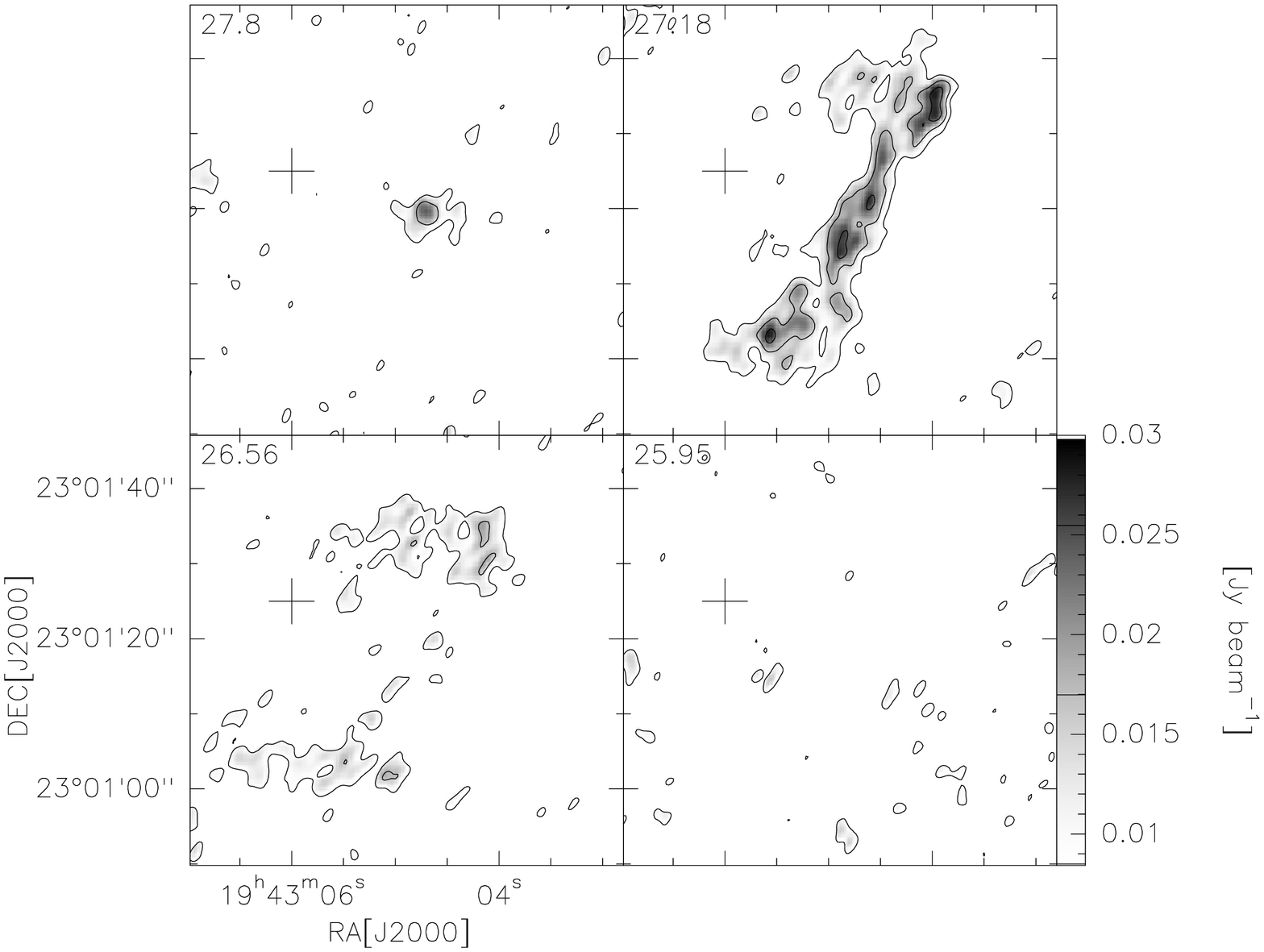}
\vspace{0.2cm}
 \caption{
{\bf V27}. Same as Figure~\ref{fig:v11ch} for V27. }
\label{fig:v27ch}
\end{figure*}

%
 \begin{figure}
 \centering
 \includegraphics[width=7cm]{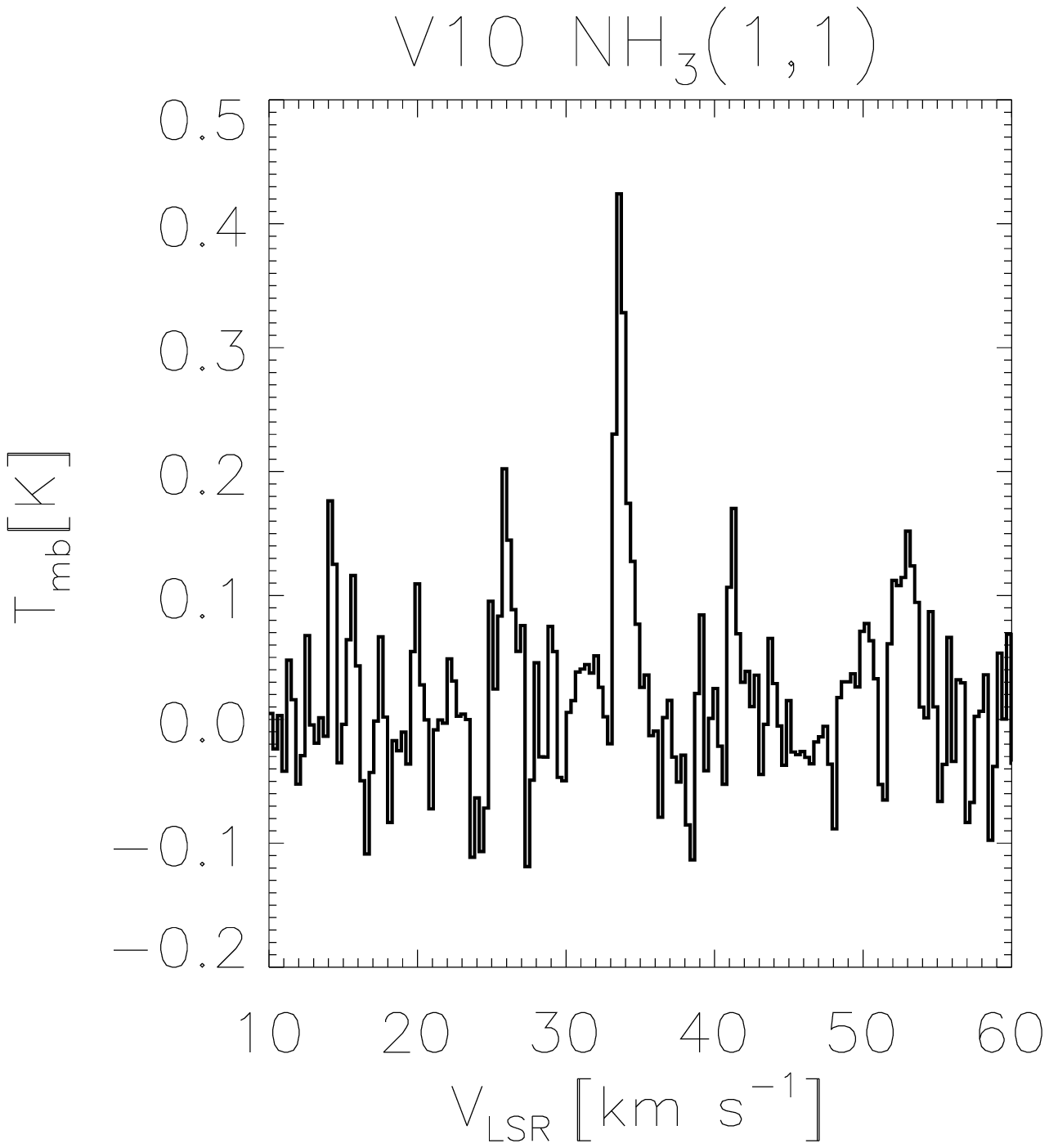}
 \caption{
{\bf V10}. NH$_3$(1,1) spectrum toward V10 taken with the Effelsberg 100-m telescope.
The (2,2) emission was not detected.
}
\label{fig:v10highres}
\end{figure}

In order to derive the rotation temperature (Section~\ref{sec:phys}) we also need 
to convert the VLA spectra in units of temperature.
The conversion from flux per beam to brightness temperature measured
in the synthesized beam, $T_{\rm sb}$, is obtained using the relation (at the frequency of
the NH$_3$(1,1) line):
\begin{equation}
%
T_{\rm sb} {\rm [K]} = 2.18 
\frac{  F_\nu {\rm [mJy/beam]}  }
{ \theta_{\rm min}{\rm [\arcsec]} \, \theta_{\rm max}{\rm [\arcsec]}   }
\label{eq:Tsb}
\end{equation}
where $\theta_{\rm min}$ and $\theta_{\rm max}$ indicate respectively the minor and major
axes at half power of the elliptical synthesized beam.  

%
 \begin{figure}
 \centering
 \includegraphics[width=7cm]{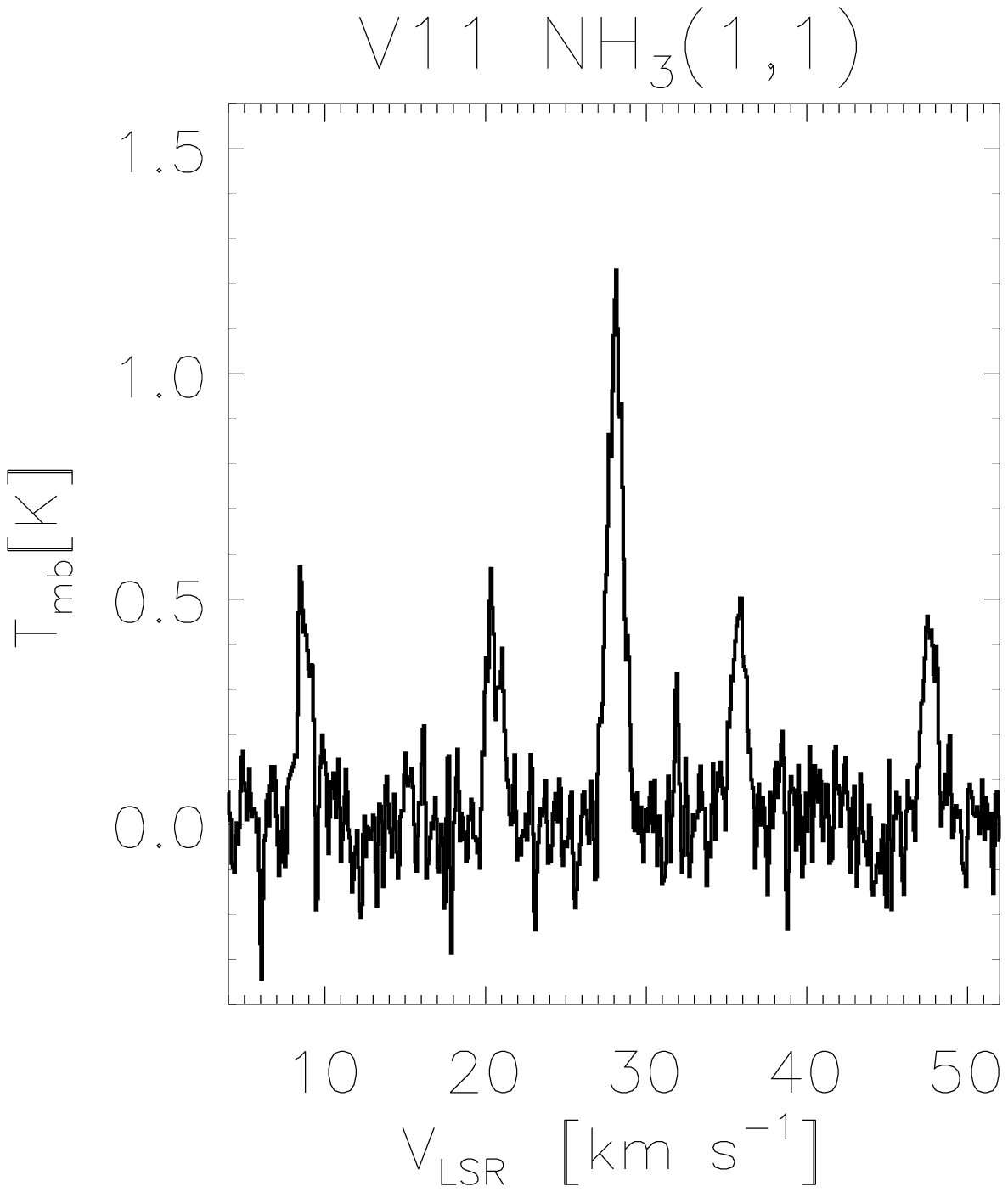}
 \includegraphics[width=7cm]{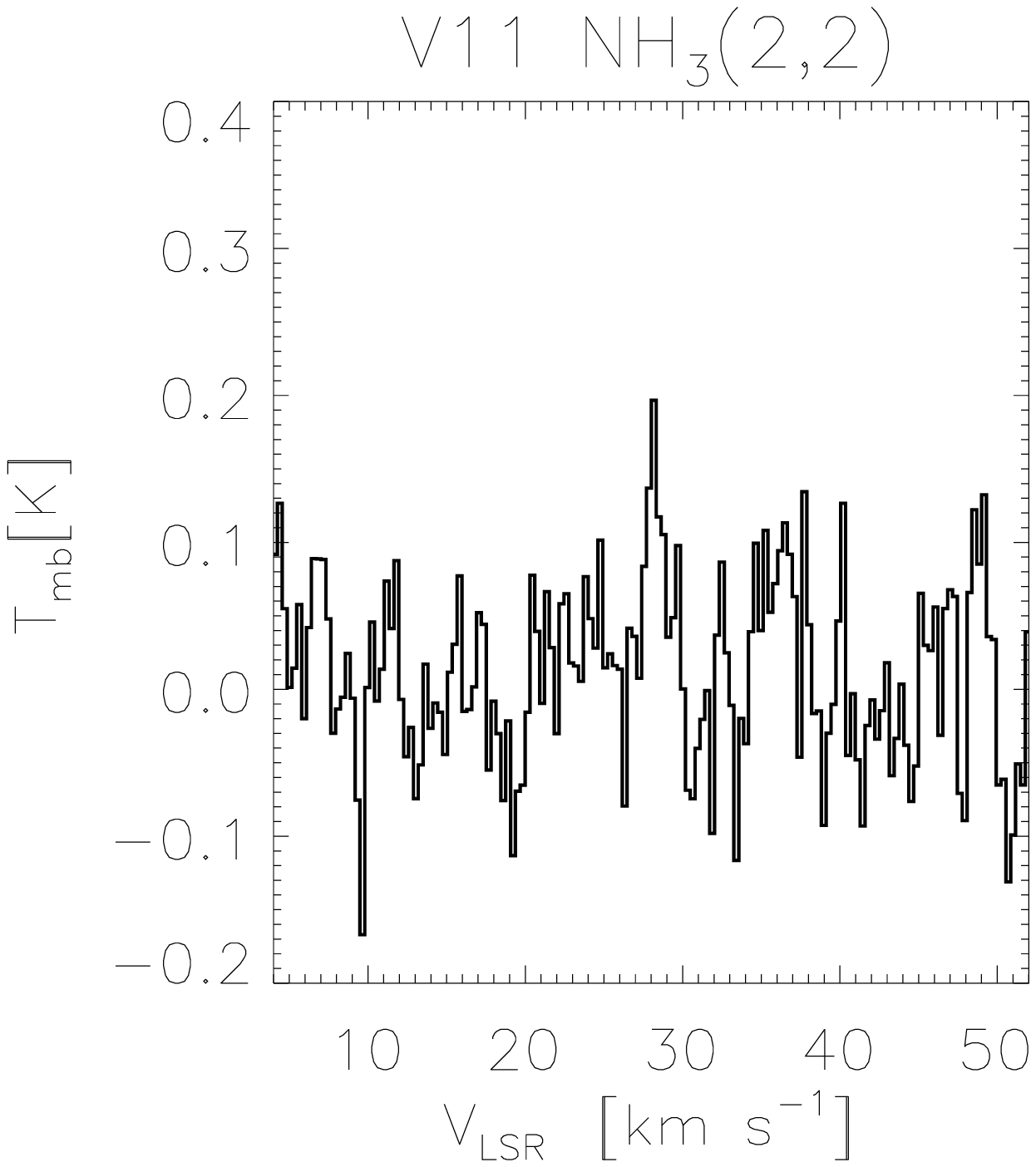}
 \caption{
{\bf V11}. {\it Top.} NH$_3$(1,1) spectrum toward V11 taken with the Effelsberg 100-m telescope.
{\it Bottom.} NH$_3$(2,2)  spectrum toward V11 smoothed at a velocity resolution of 0.3\,km\,s$^{-1}$.
}
\label{fig:v11highres}
\end{figure}

\subsection{Effelsberg 100-m Telescope}
\label{sec:100m}

Single-point spectra of the four BLAST cores were also obtained with the
MPIfR 100-m telescope and its 1.3-cm primary focus receiver.
The pointing positions used for the 100-m telescope were different from those
used with the VLA and are shown as asterisks in Figures~\ref{fig:v10} to \ref{fig:v33}.
Also shown in these figures is the contour representing the half power width of the 100-m beam.
The NH$_3$(1,1) and (2,2) lines were observed simultaneously within the same band,
in position-switched mode (with a beam-throw of $200\,\arcsec$ along the RA direction), with a
velocity resolution of $0.077 \,$km\,s$^{-1}$ and $\sim 10$\,minutes of on-source
integration time.

%
 \begin{figure}
 \centering
 \includegraphics[width=7cm]{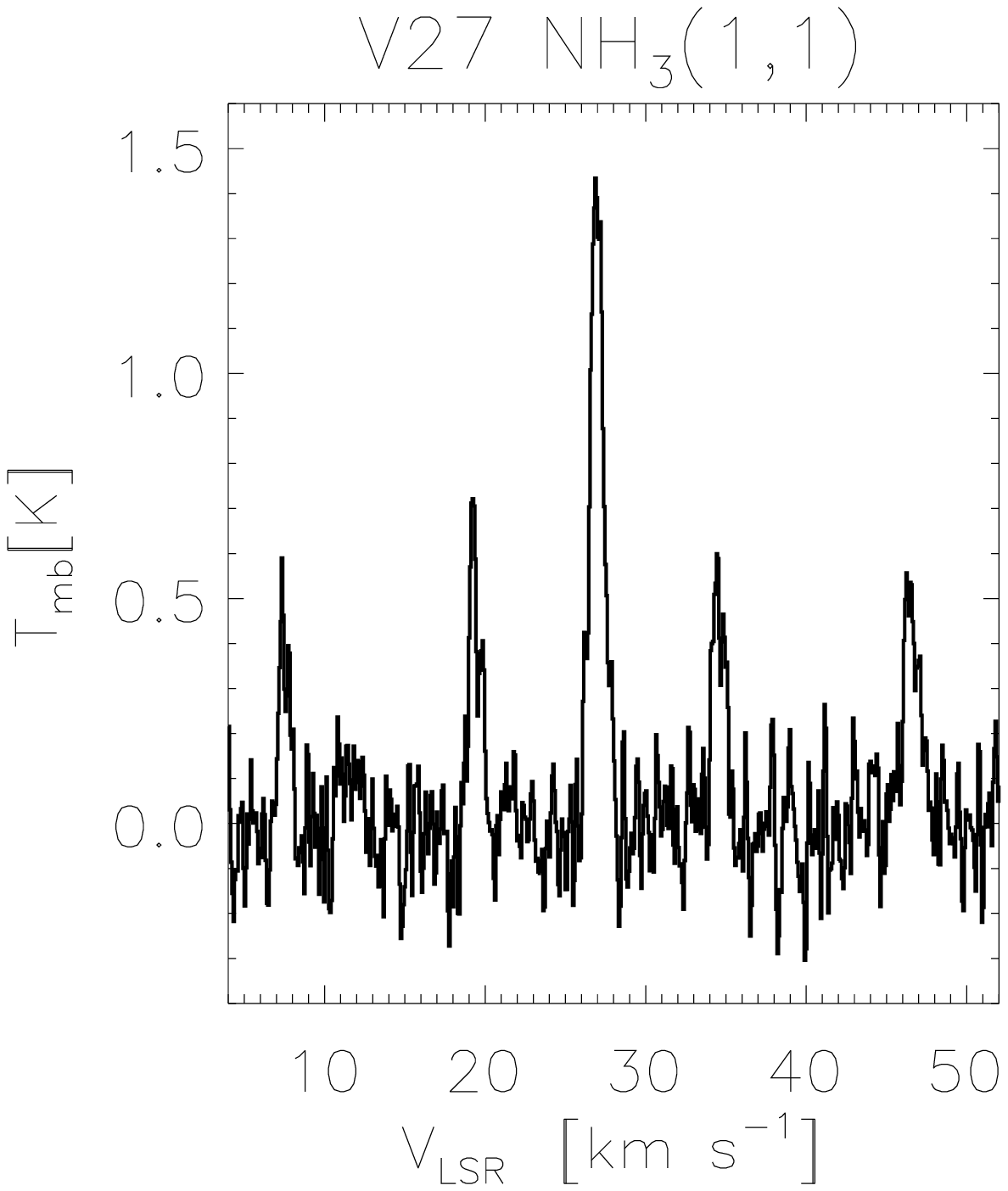}
 \includegraphics[width=7cm]{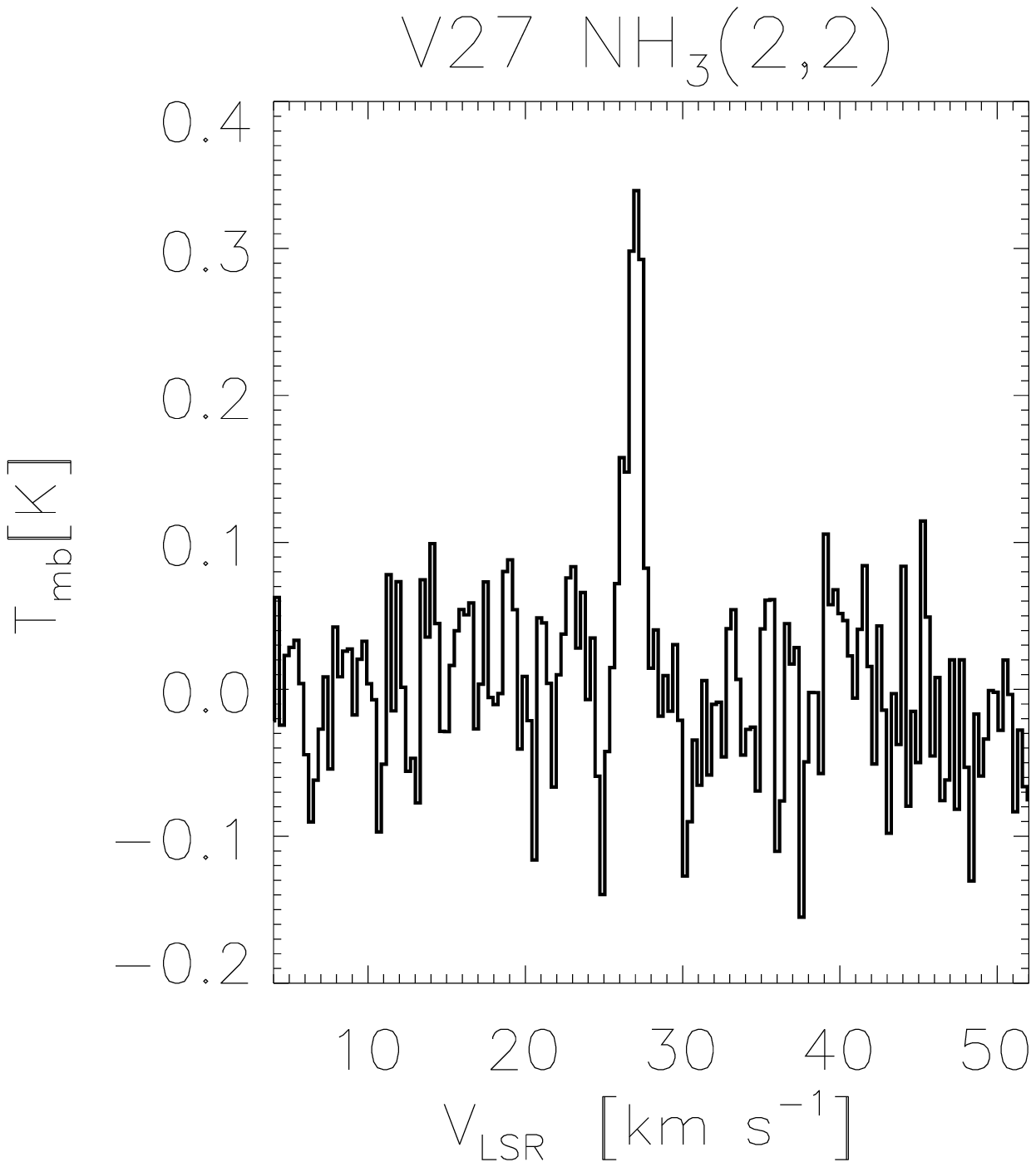}
 \caption{
{\bf V27}. Same as Figure~\ref{fig:v11highres} for V27.
}
\label{fig:v27highres}
\end{figure}

The calibration of the 100-m data followed  a standard
procedure\footnote{ { http://www.mpifr-bonn.mpg.de/div/effelsberg/ \\ 
calibration/calib.html} }
where the arbitrary noise tube units were converted to the antenna temperature scale,
and both an atmospheric opacity and elevation corrections have been applied to
the antenna gain\footnote{ { http://www.mpifr-bonn.mpg.de/staff/tpillai/eff\_calib/eff\_calib.html}  }.
The main beam efficiency was $\eta_{\rm mb}=0.52$, the FWHM of the 100-m beam was 39\,\arcsec
and the resulting noise RMS in the spectra was $\sim 100\,$mK,
corresponding to $\sim 60\,$mJy\,beam$^{-1}$ (at the native spectral resolution of $0.077 \,$km\,s$^{-1}$).
%

For comparison with the VLA data (Section~\ref{sec:100mspec}) the 100-m spectra were
also converted to Jy using the conversion formula:
\begin{equation}
S_{\rm \lambda}[{\rm Jy}] =
2.65 \left ( \frac{ \theta_{\rm mb}[{\rm arcmin}] } { \lambda[{\rm cm}] } \right )^2
T_{\rm mb}
\end{equation}
where $\theta_{\rm mb}$ and $T_{\rm mb}$ are the main-beam FWHM and brightness temperature, respectively.

\subsection{VLA Maps and Spectra}
\label{sec:vlamap}

As mentioned earlier, the VLA-D has been used to  map
the NH$_3$(1,1) and (2,2) inversion transitions. The corresponding maps and
integrated spectra (the integration area is represented by the dot-dashed boxes)
are shown in Figures~\ref{fig:v10} to \ref{fig:v33}.
Figures~\ref{fig:v11ch} and \ref{fig:v27ch} show channel maps of the NH$_3$(1,1)
main line toward sources V11 and V27.
The NH$_3$(1,1) line has been detected toward all four BLAST cores,whereas the (2,2) transition has been detected in V11, V27 and V33.
This is an excellent result, also because  the positional errors of the BLAST sources in Vulpecula,
which were estimated by \citet{chapin2008} to be $\sim 7 - 70\,\arcsec$, could make these detections more difficult.

%
 \begin{figure}
 \centering
 \includegraphics[width=7cm]{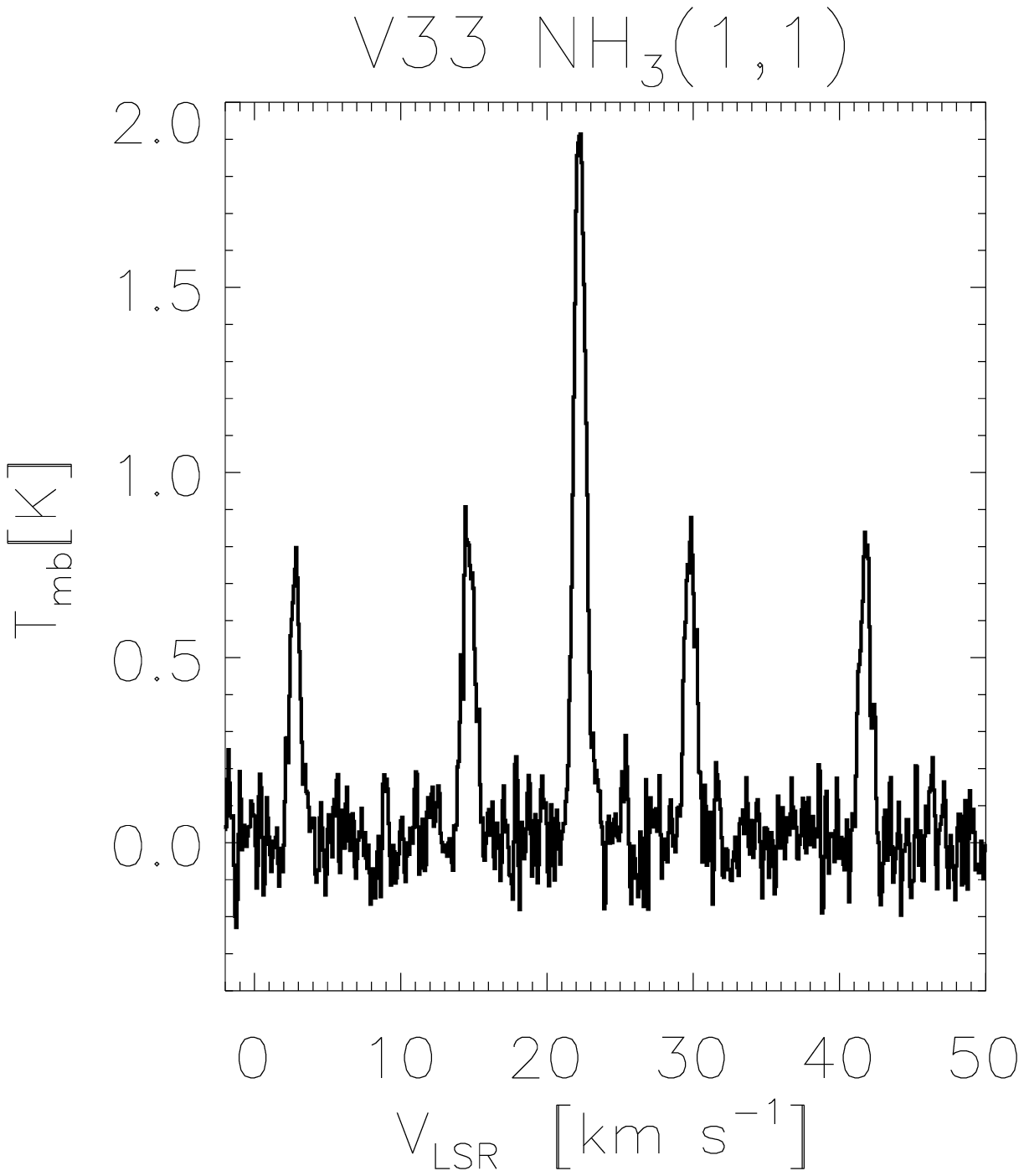}
 \includegraphics[width=7cm]{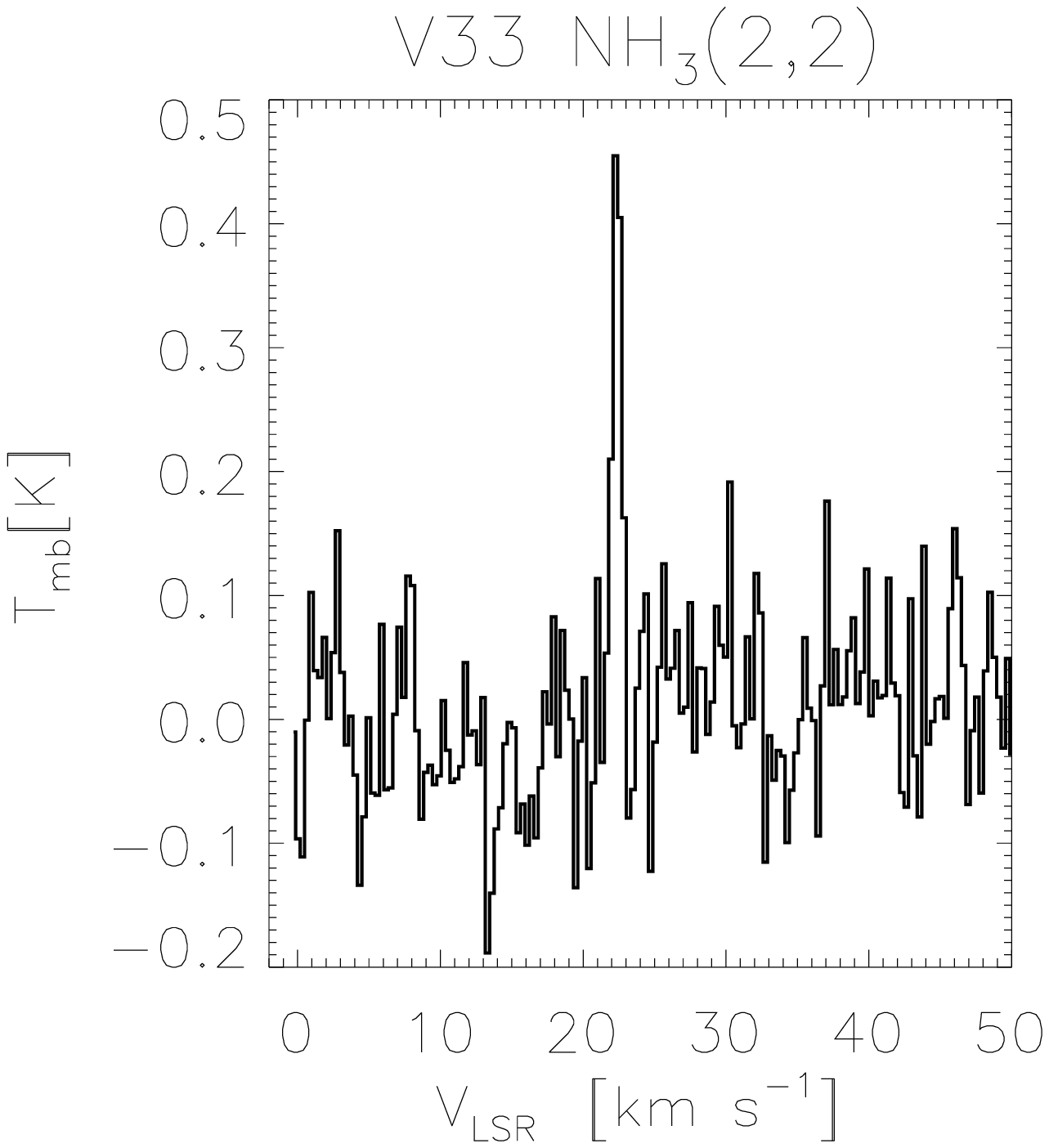}
 \caption{
{\bf V33}. Same as Figure~\ref{fig:v11highres} for V33.
}
\label{fig:v33highres}
\end{figure}

%
 \begin{figure}
 \centering
 \includegraphics[width=7cm]{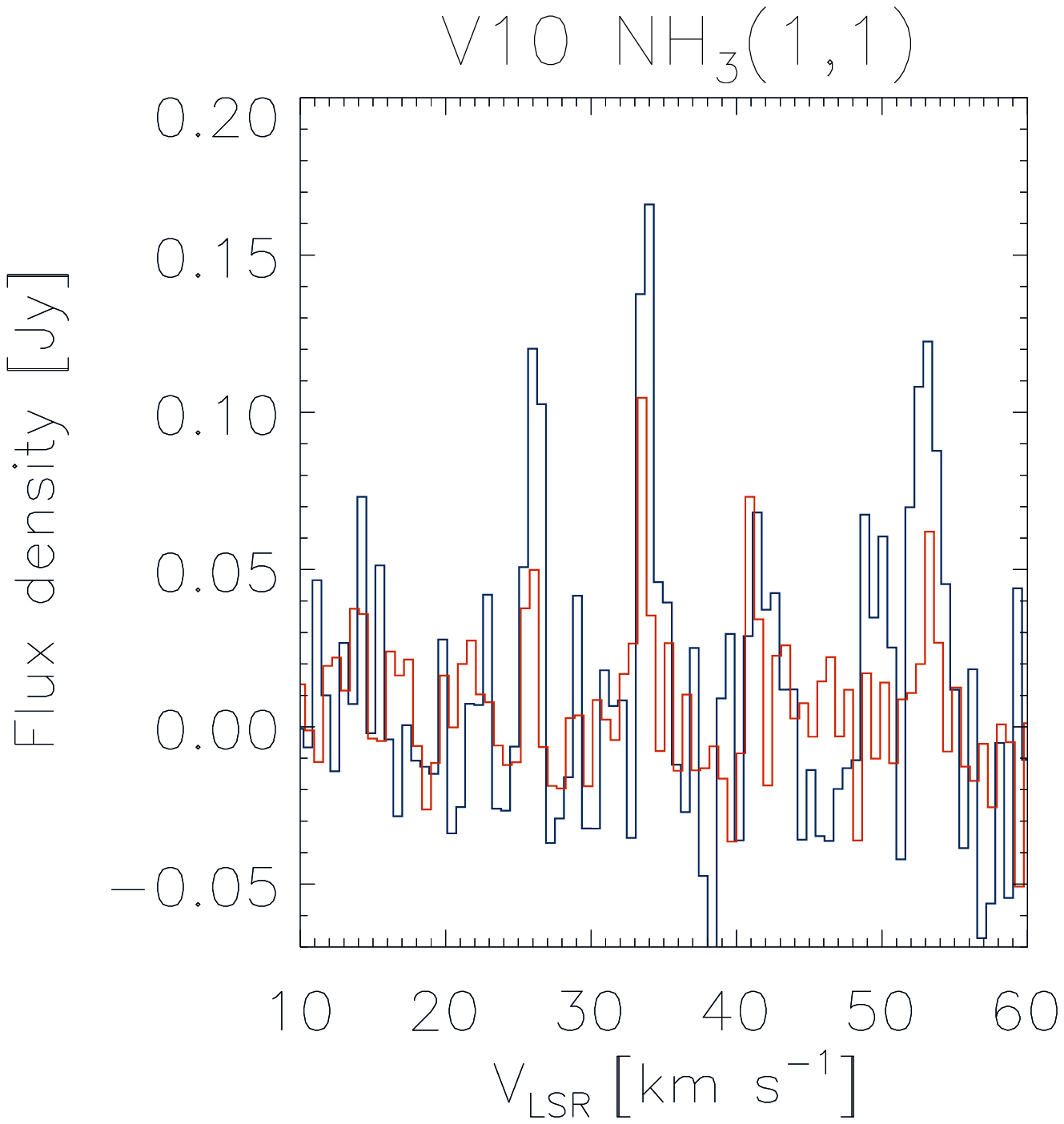}
 \caption{
{\bf V10}. Overlay of the NH$_3$(1,1) spectra of Effelsberg (blue line) and VLA (red line) in V10.
   }
\label{fig:v10overlay}
\end{figure}

We note that in V33 a reliable image of the NH$_3$(2,2) emission could not be
produced because of the low signal-to-noise ratio (SNR) and extended emission.
However, integration of the emission in
a box covering most of the NH$_3$(1,1) emission yielded a tentative detection of the (2,2) line,
as shown in the top-right panel of Figure~\ref{fig:v33}.
The hyperfine components have been detected only in the (1,1) line.
One can note that, in general, the emission appears to arise from a clumpy and filamentary
structure. In V11 and V27 this morphology can be seen even in the weaker (2,2) transition, though
at the 2-$\sigma$ level.

%
%
\begin{deluxetable}{lccccccccc}
\tablewidth{0pt}
\small
\tablecaption{Velocity, Linewidth and Total Optical Depth from Method NH$_3$(1,1).
 \label{tab:methnh3}}
\tablehead{
\colhead{} &
\multicolumn{4}{c}{{\bf VLA}$^{\rm a}$} &
\colhead{} &
\multicolumn{4}{c}{{\bf 100-m}} \\
\colhead{Source} &
\colhead{$T_{\rm ex}$} &
\colhead{$V_{\rm lsr}$} &
\colhead{$\Delta V$} &
\colhead{$\tau$} &
\colhead{} &
\colhead{$T_{\rm ex}$} &
\colhead{$V_{\rm lsr}$} &
\colhead{$\Delta V$} &
\colhead{$\tau$} \\
\colhead{} &
\colhead{K} &
\colhead{km\,s$^{-1}$} &
\colhead{km\,s$^{-1}$} &
\colhead{} &
\colhead{} &
\colhead{K} &
\colhead{km\,s$^{-1}$} &
\colhead{km\,s$^{-1}$} &
\colhead{}
}
\startdata
V10   & $4.2\pm1.2$   & $33.52\pm 0.07$  & $ <1.0$  & $ 1.8\pm 1.0$  &  & $3.23\pm0.25$  & $33.69\pm 0.02$  & $ 0.41\pm 0.05$  & $ 3.0\pm 0.9$ \\      
V11   & $5.0\pm1.1$   & $28.24\pm 0.04$  & $ <1.0$  & $ 2.0\pm 0.7$  &  & $3.97\pm0.27$  & $28.12\pm 0.01$  & $ 0.83\pm 0.03$  & $ 2.1\pm 0.3$ \\
V27   & $5.5\pm2.2$   & $27.00\pm 0.03$  & $ <1.0$  & $ 0.7\pm 0.5$  &  & $4.63\pm0.40$  & $26.98\pm 0.01$  & $ 0.70\pm 0.03$  & $ 1.6\pm 0.2$ \\
V33   & $4.3\pm0.4$   & $22.21\pm 0.03$  & $ <1.0$  & $ 2.5\pm 0.4$  &  & $4.97\pm0.25$  & $22.26\pm 0.01$  & $ 0.66\pm 0.02$  & $ 2.2\pm 0.2$ \\
\enddata
%
%
\tablenotetext{a}{The VLA line parameters represent average values, estimated by integrating
the NH$_3$(1,1) and (2,2) line emission in the dot-dashed boxes shown in Figures~\ref{fig:v10} to \ref{fig:v33}.}
\end{deluxetable}

The velocity scale shown in the spectra of Figures~\ref{fig:v10} to \ref{fig:v33} (and also
in all subsequent spectra) is referred to the NH$_3$(1,1) and (2,2) main lines.
We note the excellent agreement  with the velocities estimated by \citet{chapin2008} using $^{13}$CO$(1-0)$.
All spectral lines are clearly very narrow and the VLA observations do not resolve the lines
(see later Section~\ref{sec:phys} and Tables~\ref{tab:methnh3} and \ref{tab:gauss22}).
In the case of the (1,1) maps of sources V11 and V27  one can note in Figures~\ref{fig:v11ch}
and \ref{fig:v27ch} the sharp change in the
brightness spatial distribution in adjacent velocity channels. This indicates
a  significant velocity substructure within $\sim 1\,$km\,s$^{-1}$. In V11 and V27
these features are unlikely to be a consequence of missing flux, since
with the VLA-D we recover most of the flux
measured by the 100-m telescope (see Section~\ref{sec:100mspec}).

Although the morphological difference between these sources could be caused by both intrinsic
(e.g., geometrical effects and/or different evolutionary phases) and observational effects,
all these cores present very similar features. In fact, they all show an internal structure,
with both smaller cores, or fragments,
 and an inter-core emission, which also appears to be  ``filamentary''.
The fragments inside the cores have sizes $\la 0.05\,$pc and we are unable to resolve structures
smaller than $\simeq 0.034\,$pc (7100\,AU).

\subsection{Effelsberg Spectra}
\label{sec:100mspec}

The Effelsberg telescope was used to observe the four BLAST sources toward their central
positions at higher spectral resolution. The resulting spectra are shown in Figures~\ref{fig:v10highres}
to \ref{fig:v33highres},
where we note that the hyperfine components of both the inner and outer satellites are partly resolved,
an indication of the narrow linewidths in this source.

\begin{figure}
 \centering
 \includegraphics[width=7cm]{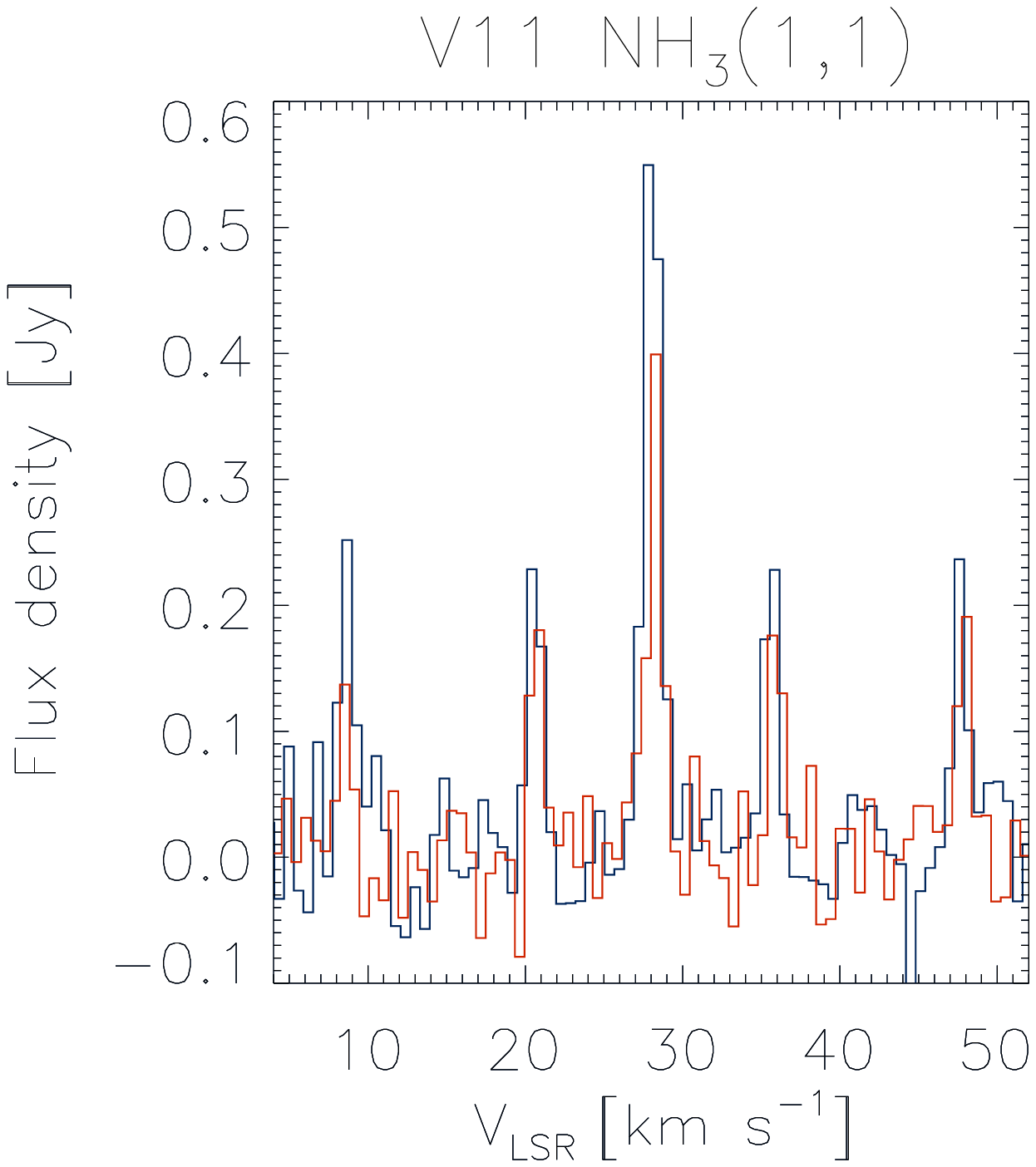}
 \includegraphics[width=7cm]{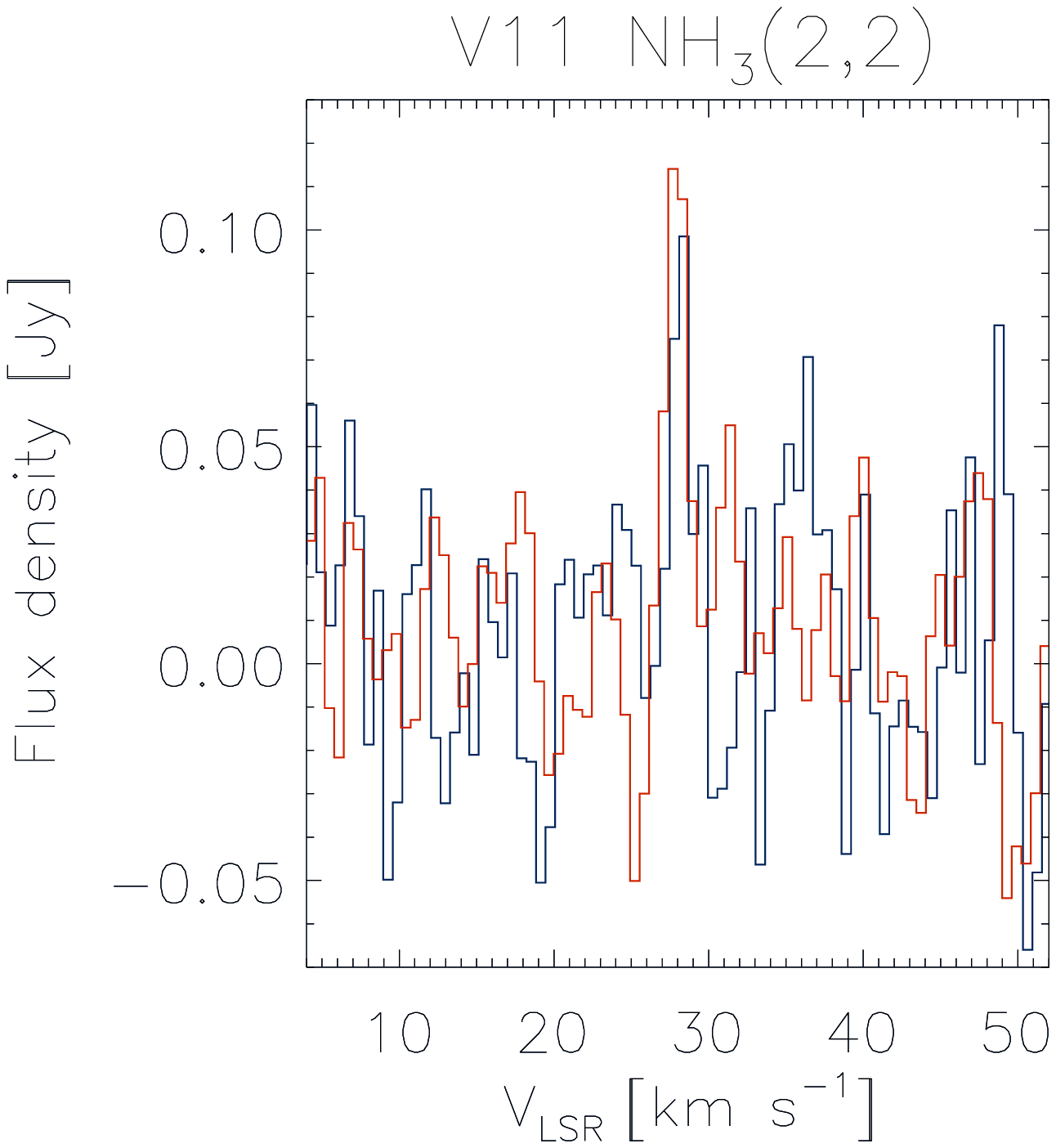}
 \caption{
{\bf V11}. {\it Top.} Overlay of the NH$_3$(1,1) spectra of Effelsberg
(blue line) and VLA (red line) in V11.
{\it Bottom.} Same as above for NH$_3$(2,2).
   }
\label{fig:v11overlay}
\end{figure}

\begin{figure}
 \centering
 \includegraphics[width=7cm]{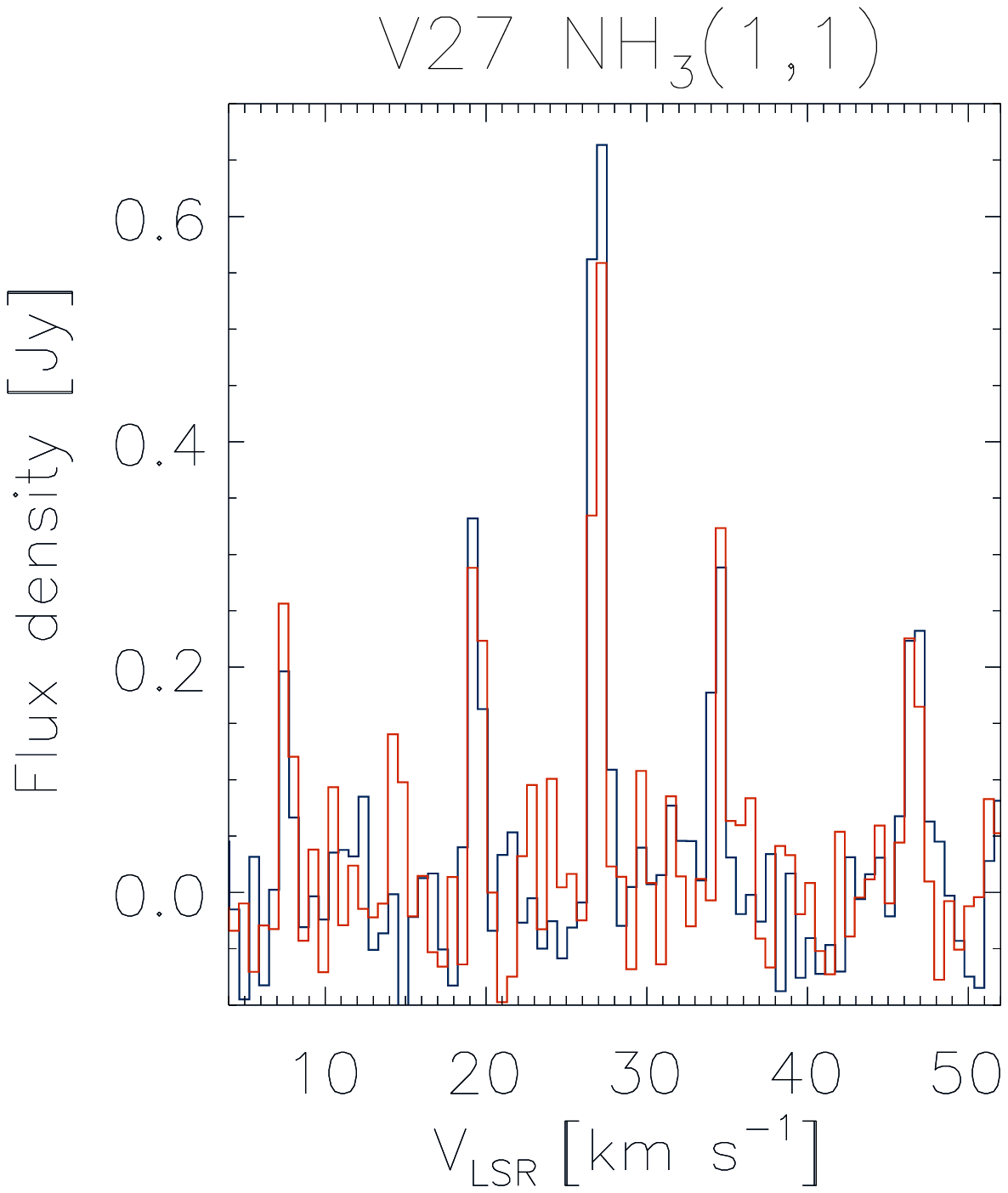}
 \includegraphics[width=7cm]{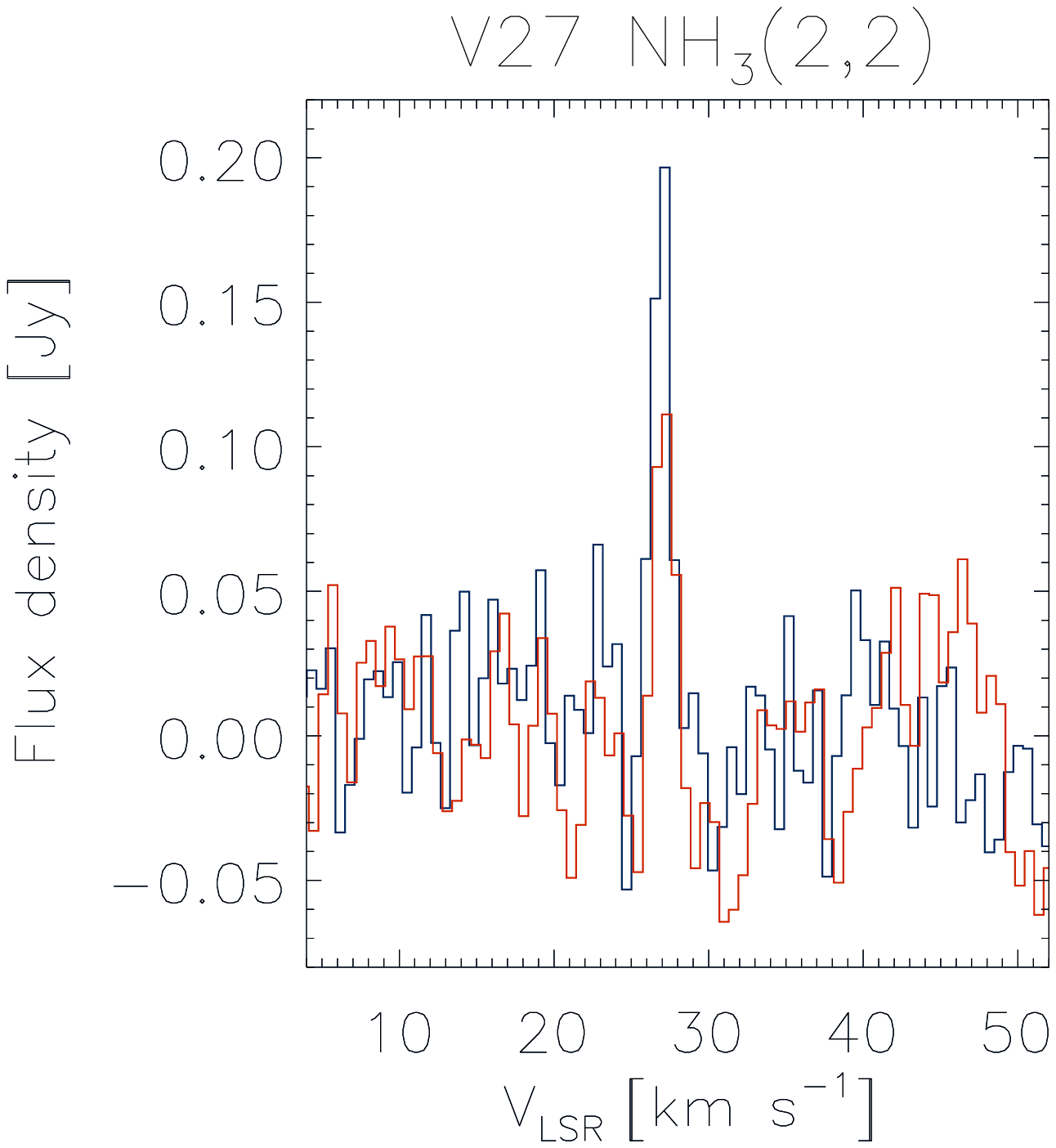}
 \caption{
{\bf V27}. Same as Figure~\ref{fig:v11overlay} for V27.
   }
\label{fig:v27overlay}
\end{figure}

For comparison with the VLA, the spectra obtained with the 100-m telescope
were resampled at the same spectral resolution as the VLA, and the results are
shown in Figures~\ref{fig:v10overlay} to \ref{fig:v33overlay}. The flux measured by the VLA
is almost always (exceptions are the NH$_3$(2,2) spectra toward V11 and V33) 
lower than that observed with the 100-m telescope, probably as a result of both
low SNR in the VLA spectra and extended emission that is filtered out by the interferometer.
If we compare the main-line emission, the amount of the single-dish flux lost by the VLA
varies from $\simeq 20-30\,$\% in V27 and V11 to $\simeq 50\,$\% in V10 and V33.
In V11 and V33 the VLA recovers most of the NH$_3$(2,2) single-dish flux.

\section{ANALYSIS}
\label{sec:analysis}

\subsection{Derivation of Physical Parameters with METHOD NH$_3$(1,1)}
\label{sec:phys}

The NH$_3$(1,1) and (2,2) inversion lines show an electric quadrupole and magnetic
hyperfine structure (see \citealp{ho1983} for a review).
The (1,1) spectra were fitted using a non-linear least-square method
(METHOD NH$_3$(1,1) of the CLASS program) which takes into account the 18 hyperfine components.
This method can determine the optical depths and linewidths assuming that all components
have equal excitation temperatures, that the line separation is
fixed at the laboratory value and that the linewidths are identical.

The NH$_3$(2,2) lines are weaker and the sensitivity of these observations is less 
than the sensitivity of the NH$_3$(1,1) data. Thus, the quadrupole hyperfine structure 
of the NH$_3$(2,2) transition is not visible when the emission is averaged
over the same large area as the (1,1) line, 
shown by the dot-dashed boxes in Figures~\ref{fig:v10} to \ref{fig:v33}.
The hyperfine structure of the (2,2) line becomes visible in sources V11 and V27 (see Figure~\ref{fig:22mask})
only when the emission is averaged on smaller sub-regions, such as the 
ones shown by the boldface dashed contours in Figures~\ref{fig:v11} and \ref{fig:v27}.
Therefore, for the purpose of using METHOD NH$_3$(1,1) of the CLASS program we always 
fitted the (2,2) profiles (obtained by integrating the VLA maps in the
same region as the (1,1) transition) by simple Gaussians.
Due to the hidden magnetic hyperfine structure, this tends to overestimate
the intrinsic linewidths \citep{bach1987}.

The results of the NH$_3$(1,1) fitting are listed in Table~\ref{tab:methnh3}.
The 100-m spectra have the advantage of a better SNR and spectral resolution
and therefore the derived parameters are more accurate, though they do not
necessarily describe the same volume of gas. The effect of the better spectral
resolution is clearly visible in the resulting linewidths (see Table~\ref{tab:methnh3}). 
As mentioned earlier the VLA does not resolve the lines and thus the VLA linewidths should 
be interpreted only as upper limits. On the other hand
the 100-m results show that the linewidths are indeed different and very narrow.
A temperature of $T_{\rm k}=14\,$K corresponds to a NH$_3$ thermal linewidth
of $\Delta V_{\rm th} = 0.19\,$km\,s$^{-1}$
($\Delta V_{\rm th} = \sqrt{8 \, \ln 2 \, k T_{\rm k} / (17 \, m_{\rm H}) } $, where $k$ is the
Boltzmann constant and $m_{\rm H}$ is the atomic hydrogen  mass). 
Therefore, non-thermal motions definitely give an important contribution in these sources.
The overall linewidths, however, are smaller (or much smaller in the case of V10)
than those previously observed toward (warmer) HMPOs by, e.g., \citet{mol96}, who found median 
linewidths $\simeq 1.7 - 1.8$\,km\,s$^{-1}$, and also by \citet{motte2007} who found linewidths
$\ga 2.0$\,km\,s$^{-1}$. 

%

%
%
\begin{deluxetable}{lccccccc}
\tablewidth{0pt}
\small
\tablecaption{Results of Gaussian Fits to the NH$_3$(2,2) Lines.
 \label{tab:gauss22}}
\tablehead{
\colhead{} &
\multicolumn{3}{c}{{\bf VLA}} &
\colhead{} &
\multicolumn{3}{c}{{\bf 100-m}} \\
\colhead{Source} &
\colhead{$T_{\rm sb}$} &
\colhead{$V_{\rm lsr}$} &
\colhead{$\Delta V$} &
\colhead{} &
\colhead{$T_{\rm mb}$ } &
\colhead{$V_{\rm lsr}$} &
\colhead{$\Delta V$} \\
\colhead{} &
\colhead{K} &
\colhead{km\,s$^{-1}$} &
\colhead{km\,s$^{-1}$} &
\colhead{} &
\colhead{K} &
\colhead{km\,s$^{-1}$} &
\colhead{km\,s$^{-1}$}
}
\startdata
V11    & $0.46\pm 0.09$   & $27.91\pm 0.13$    & $1.4\pm  0.3$   &  & $0.16\pm 0.06$   & $28.23\pm 0.19$     & $1.44\pm 0
.56$  \\
V27    & $0.24\pm 0.06$   & $27.13\pm 0.15$    & $1.3\pm  0.3$   &  & $0.33\pm 0.06$   & $26.96\pm 0.08$     & $1.23\pm 0
.18$  \\
V33    & $0.62\pm 0.14$   & $22.27\pm 0.13$    & $< 1.0$         &  & $0.52\pm 0.09$   & $22.36\pm 0.04$     & $0.72\pm 0
.08$  \\
\enddata
\tablecomments{
The VLA fits have been performed on the integrated spectra in the area shown by the dot-dashed boxes
in Figures~\ref{fig:v11} to \ref{fig:v33}.
}
\end{deluxetable}

Source V11 has the largest linewidth,
as measured by both (1,1) and (2,2) lines. We also note that V11 is the only source where the
velocity structure shown in Figure~\ref{fig:v11ch} is suggesting the presence of either
two cores at separate velocities or a velocity gradient, positive S to N,
of $\simeq 11\,$km\,s$^{-1}$\,pc$^{-1}$. Neither of these can be ruled out on the basis
of the present data. Incidentally, V11 is also the only source with a clear positional
association with a likely protostar (see Section~\ref{sec:evo}).

\begin{figure}
 \centering
 \includegraphics[width=7cm]{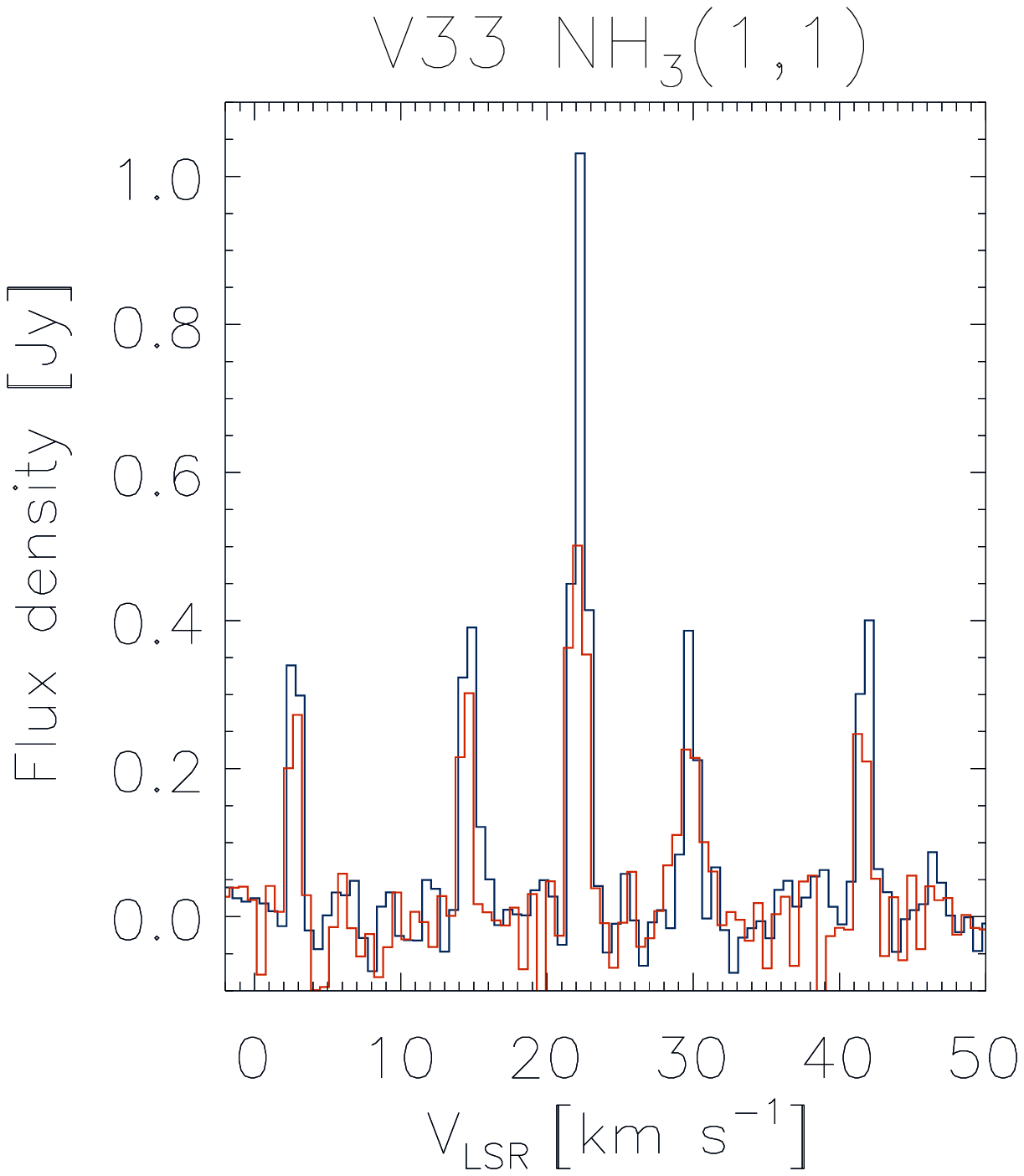}
 \includegraphics[width=7cm]{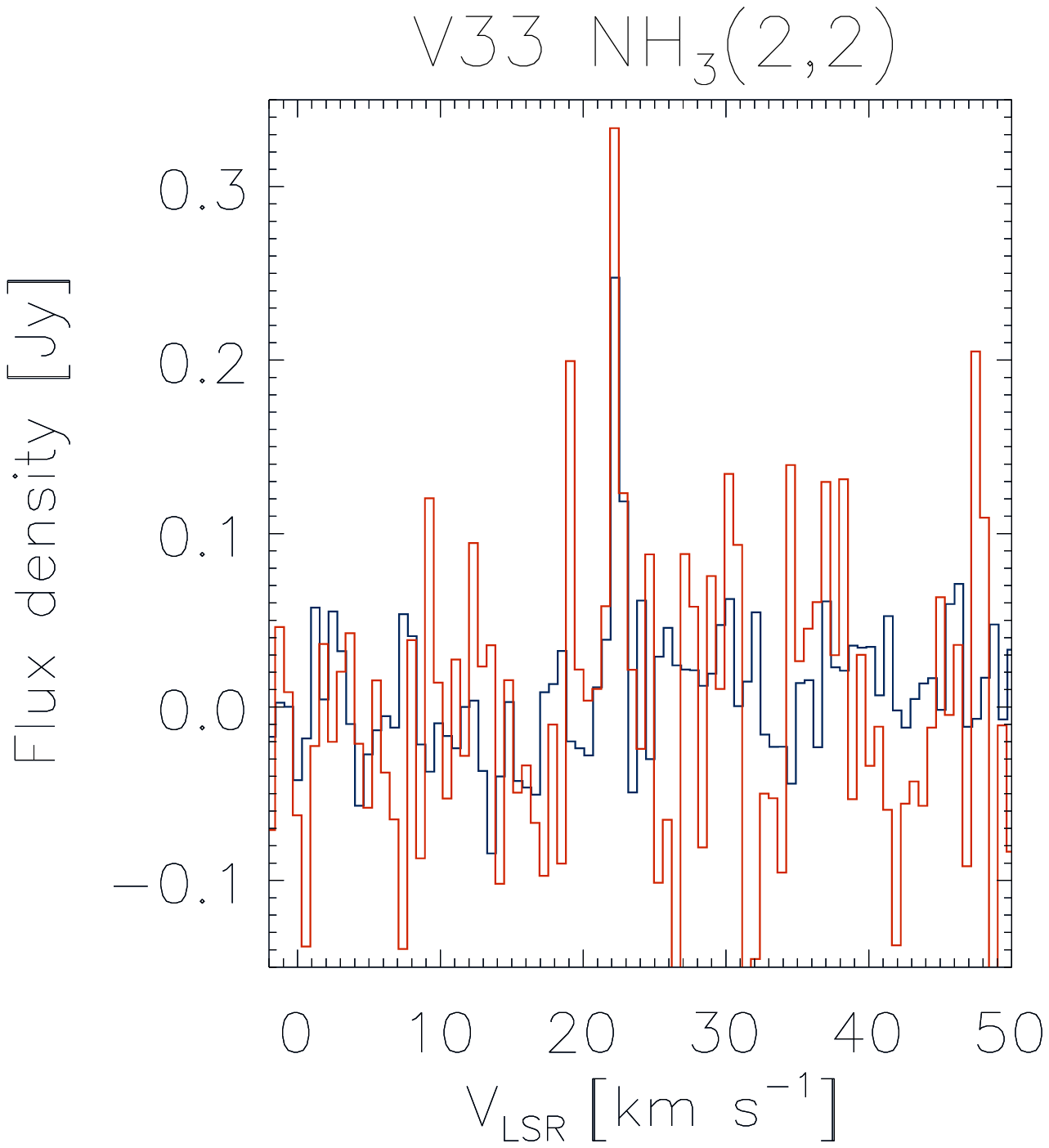}
 \caption{
{\bf V33}. Same as Figure~\ref{fig:v11overlay} for V33.
   }
\label{fig:v33overlay}
\end{figure}

The results of the Gaussian fits to the NH$_3$(2,2) profiles are listed in Table~\ref{tab:gauss22}.
Because the Gaussian fit does not take into account the hyperfine structure of the line, we
cannot compare directly the linewidths in Tables~\ref{tab:methnh3} and \ref{tab:gauss22}.
We also note the same basic behaviour among those sources with a (2,2) detection: source
V33 has in fact the smallest linewidth as measured by both (1,1) and (2,2) lines.
However, because in the VLA spectrum the (2,2) line is barely two channels wide
(see Figure~\ref{fig:v33overlay}) in Table~\ref{tab:gauss22} we only list an upper limit.


%

%
%
\begin{deluxetable}{lccccccc}
\tablewidth{0pt}
\small
\tablecaption{Temperature and Column Density
 \label{tab:rottemp}}
\tablehead{
\colhead{} &
\multicolumn{3}{c}{{\bf VLA}$^{\rm a}$} &
\colhead{} &
\multicolumn{3}{c}{{\bf 100-m}} \\
\colhead{Source} &
\colhead{$T_{12}$} &
\colhead{$T_{\rm k}$} &
\colhead{$N({\rm NH_3})$} &
\colhead{} &
\colhead{$T_{12}$} &
\colhead{$T_{\rm k}$} &
\colhead{$N({\rm NH_3})$} \\
\colhead{} &
\colhead{K} &
\colhead{K} &
\colhead{$\times 10^{14}$\,cm$^{-2}$} &
\colhead{} &
\colhead{K} &
\colhead{K} &
\colhead{$\times 10^{14}$\,cm$^{-2}$}
}
\startdata
%
%
V11    & $10.5\pm 0.5$  & $11.1\pm 0.6$  & $21.5\pm4.7$  &  & $11.9\pm 1.1$  & $12.8\pm 1.3$  & $6.6\pm 1.0$ \\
V27    & $8.7\pm 0.5$   & $9.0\pm 0.6$   & $34.1\pm8.4$  &  & $14.3\pm 0.6$  & $16.0\pm 1.0$  & $3.9\pm 0.4$ \\
V33    & $9.2\pm 0.2$   & $9.5\pm 0.2$   & $37.8\pm6.7$  &  & $12.2\pm 0.3$  & $13.1\pm 0.5$  & $6.5\pm 0.4$  \\
\enddata
\tablenotetext{a}{The VLA line parameters represent average values, estimated by integrating
the NH$_3$(1,1) within the sub-regions enclosed by the dashed contours
in Figures~\ref{fig:v11} to \ref{fig:v33} (see text and Figure~\ref{fig:v27region}).}
\end{deluxetable}

\subsection{Temperature} 
\label{sec:trot}

In V11, V27, and V33, where the (2,2) line was detected,
we were also able to determine the rotation
temperature, $T_{\rm 12}$ (and thus the kinetic temperature $T_{\rm k}$),
and the column density using the method of \citet{ung1986} and \citet{bach1987}.
In this method the required data were: {\it (i)} the product
$\tau (T_{\rm ex}- T_{\rm bg})$ for the (1,1) line, where $T_{\rm ex}$ and $T_{\rm bg}=2.725\,$K are
the excitation and background temperatures, respectively, and $\tau$ is the optical depth; 
{\it (ii)} the linewidth $\Delta V (1,1)$;
and {\it (iii)} the (2,2) integrated intensity.
Once $T_{\rm 12}$ has been estimated, the
kinetic temperature can be determined using the analytical
expression of \citet{tafalla2004}:
\begin{equation}
T_{\rm k} = \frac{T_{\rm 12}} {1 - \frac{T_{\rm 12}}{42} \ln [1 + 1.1 \exp(-\frac{16}{T_{\rm 12}} ) ] }
\label{eq:Tk}
\end{equation}
which is an empirical expression that fits the $T_{\rm 12} - T_{\rm k}$ relation obtained using
a radiative transfer model.

The results are shown in Table~\ref{tab:rottemp}.  
The VLA physical parameters represent average values, estimated by first integrating
the NH$_3$(1,1) in the smaller sub-regions enclosed by the boldface dashed contours shown
in Figures~\ref{fig:v11} to \ref{fig:v33} (see also Sect.~\ref{sec:mass}); then, the parameters
obtained in each of these sub-regions have been averaged together and the results listed 
in Table~\ref{tab:rottemp}. Because of the coarse spectral resolution of the VLA spectra, 
when evaluating the physical parameters from the VLA data we have replaced the original linewidths
with the 100m linewidths. Although the 100m linewidths represent themselves averages
over a larger source area compared to the smaller sub-regions shown in Figures~\ref{fig:v11} to \ref{fig:v33},
they give a better estimate of the $\Delta V (1,1)$ values as compared to the upper limits given by the
VLA linewidths. In addition, because of the weakness of the (2,2) transition, we could not obtain separate
(2,2) integrated spectra for the sub-regions. As a consequence, we have always used the
NH$_3$(2,2) spectrum obtained by integrating over the entire source, as shown in 
Figures~\ref{fig:v11} to \ref{fig:v33}.

The small linewidths listed in Table~\ref{tab:methnh3}
and the low kinetic temperatures of Table~\ref{tab:rottemp} again suggest small internal
motions and therefore very quiescent conditions.  The four BLAST cores appear to be even colder
and more quiescent than the HMSC candidates observed by \citet{Sridharan05}, who found
in their sample of candidate HMSCs a median linewidth of 1.5\,km\,s$^{-1}$, comparable with
that of \citet{mol96}, and a median rotation temperature of 16.9\,K.
The average temperature of our cores is also somewhat lower than the mean kinetic temperature of 15\,K
found in IRDCs by \citet{pillai2006}. 

%
%
%
\begin{deluxetable}{lcccccccccc}
\tablewidth{0pt}
\small
\tablecaption{Size, Mass and Density
 \label{tab:phys}}
\tablehead{
\colhead{Source} &
\multicolumn{2}{c}{Diameter$^{\rm a}$} &
\colhead{} &
\colhead{$M_{\rm BLAST}^{\rm b}$} &
\colhead{$M_{\rm vir}$} &
\multicolumn{2}{c}{$M_{\rm tot}^{\rm c}$} &
\colhead{} &
\multicolumn{2}{c}{$\langle n_{\rm H_2} \rangle^{\rm c}$} \\
\colhead{} &
\colhead{arcsec} &
\colhead{pc} &
\colhead{} &
\colhead{$M_\odot$} &
\colhead{$M_\odot$} &
\colhead{$M_\odot$} &
\colhead{$M_\odot$} &
\colhead{} &
\colhead{$\times 10^{5}\,$cm$^{-3}$} &
\colhead{$\times 10^{5}\,$cm$^{-3}$}
}
\startdata
%
%
V10  & 7.6   & 0.08   &  & 89   & 1.5   & -     & -     &   & -   & -    \\
V11  & 12.6  & 0.14   &  & 213  & 10.2  & 5.0   & 72.0  &   & 2.2 & 31.0 \\
V27  & 11.6  & 0.13   &  & 105  & 6.7   & 9.7   & 138.5 &   & 2.3 & 33.4 \\
V33  & 18.8  & 0.21   &  & 107  & 9.6   & 21.0  & 300.7 &   & 2.1 & 30.0 \\
\enddata
\tablecomments{
$M_{\rm tot}$ represents the total mass of the source as estimated by summing the
masses ($M_{\rm cd}$) of the sub-regions enclosed by dashed contours 
shown in Figures~\ref{fig:v11} to \ref{fig:v33}.
$\langle n_{\rm H_2} \rangle $ represents the average density among the sub-regions in which
each source has been divided.
}
\tablenotetext{a}{Estimated from the total area of the source within the 50\% contour.}
\tablenotetext{b}{BLAST core masses from \citet{chapin2008}.}
\tablenotetext{c}{The first and second value of both $M_{\rm tot}$ and 
$\langle n_{\rm H_2} \rangle $ correspond to
the values of $X[{\rm NH_3}]=10^{-7}$ and $X[{\rm NH_3}]=7\times 10^{-9}$,
respectively. The errors on $M_{\rm tot}$ are derived from the errors on $N_{\rm H_2}$ for
each separate region, and vary between $\simeq 15$ and 25\%.}
\end{deluxetable}

\subsection{Column Density and Mass}
\label{sec:mass}

The three fitting parameters described in Section~\ref{sec:trot} may also be used to
determine the column densities, $N(1,1)$ and $N(2,2)$, of the (1,1) and (2,2) lines. Then,
the total NH$_3$ column density, $N_{\rm NH_3}$, is obtained from $T_{\rm 12}$
using the method of \citet{ung1986}. 
%
The total mass in the VLA maps has been estimated by dividing each source into smaller sub-regions
(shown by the boldface dashed contours in Figures~\ref{fig:v11} to \ref{fig:v33}) and 
determining $N_{\rm NH_3}$ in each of them. The approximate size of each region 
represents a trade-off between the need to achieve a good SNR in the integrated spectra of each
region, and the need to avoid integrating over too large an area with no NH$_3$ emission.
For the VLA data, the values of temperatures and column densities averaged in these sub-regions 
in sources V11, V27 and V33 are listed in Table~\ref{tab:rottemp}

\begin{figure}
 \centering
 \includegraphics[width=8.5cm]{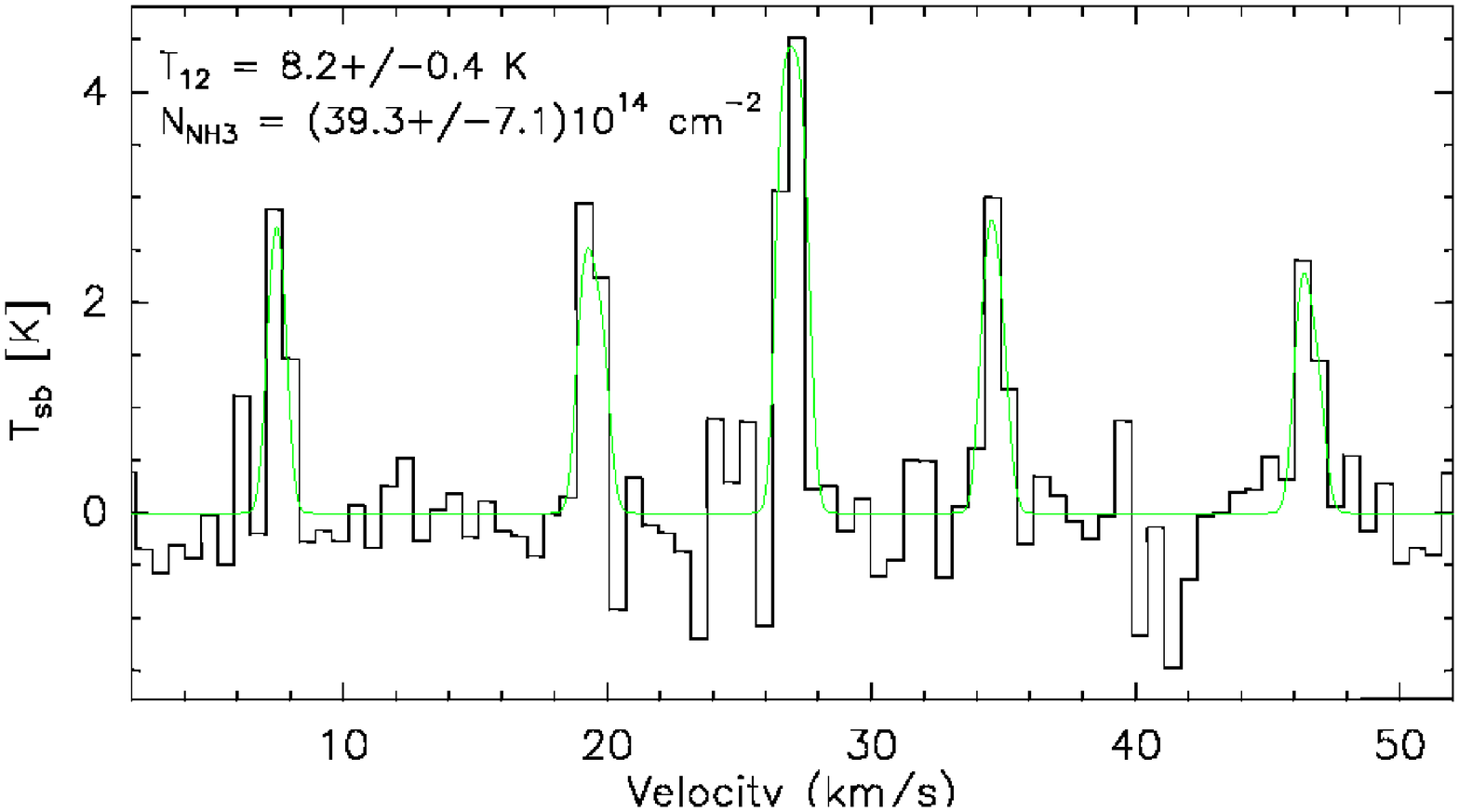}
 \includegraphics[width=8.5cm]{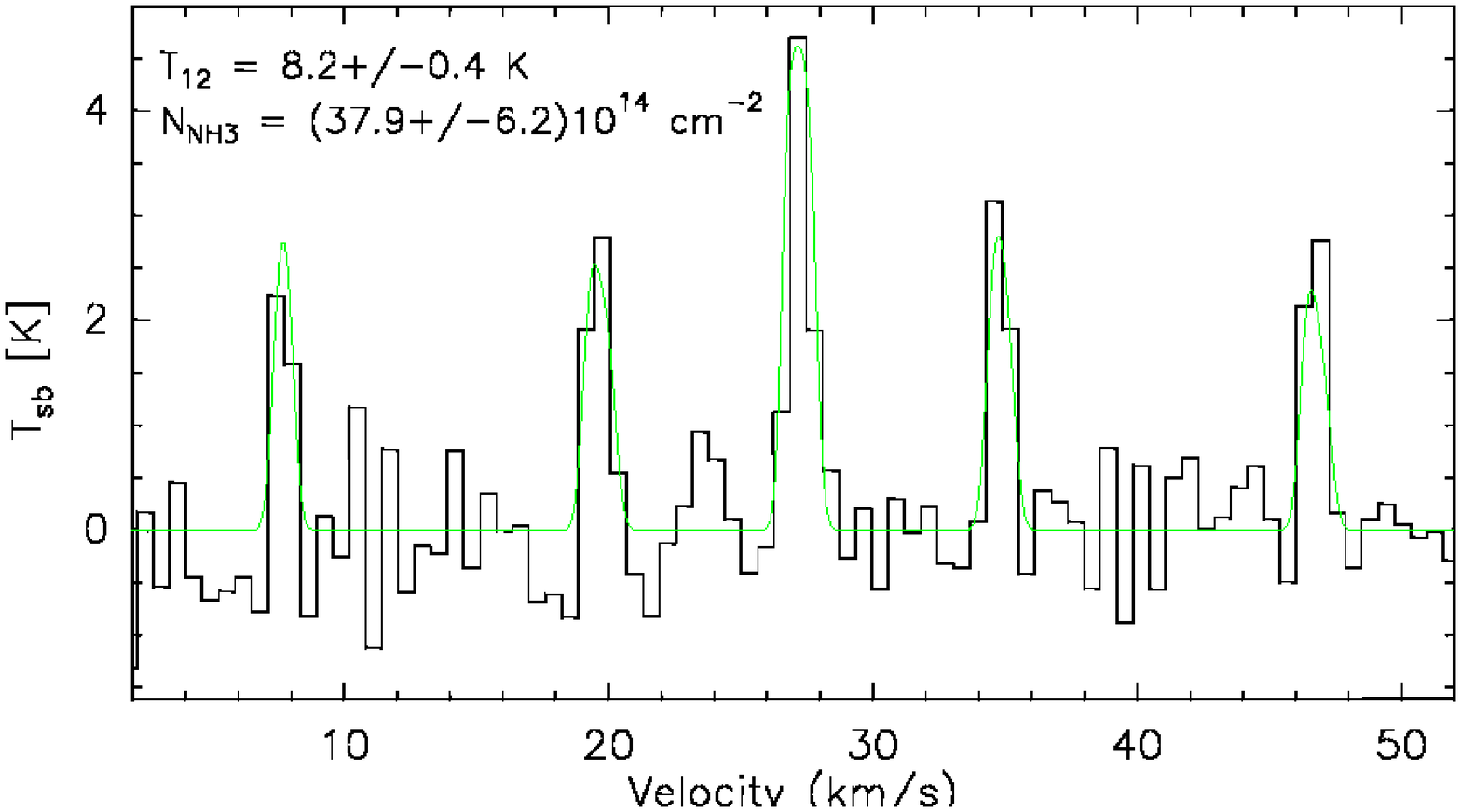}
 \includegraphics[width=8.5cm]{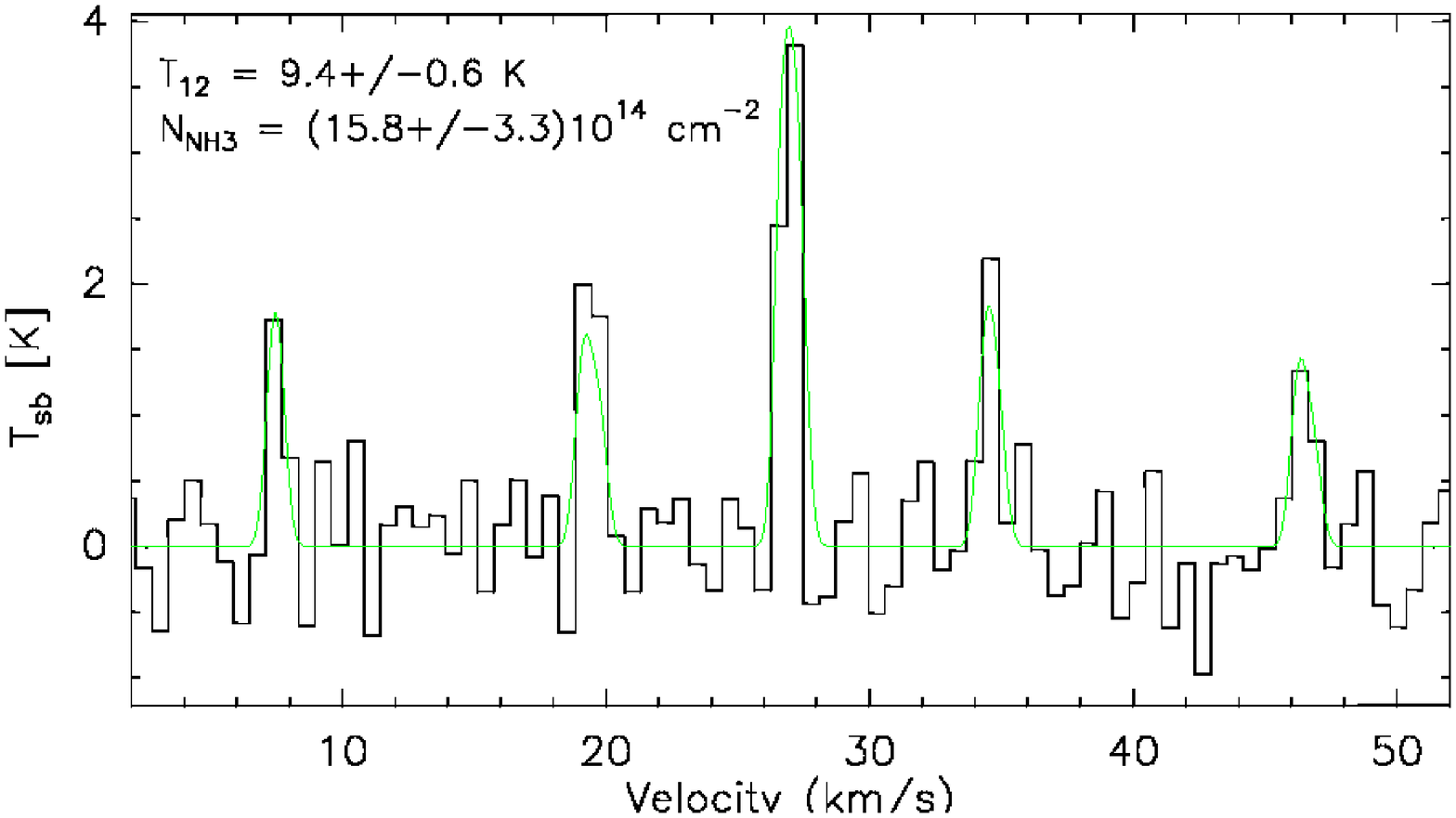}
 \caption{
VLA spectra obtained towards V27 by selecting the three smaller sub-regions 
enclosed by the boldface dashed contours shown in Figure~\ref{fig:v27}.
The solid line represents the fit obtained with METHOD NH$_3$(1,1) of the CLASS program,
by setting the VLA linewidths to the 100m values (see text).
Also shown are the values of the rotation temperature (Section~\ref{sec:trot}) 
and NH$_3$ column density as obtained in each separate sub-region 
(see Section~\ref{sec:mass} and Table~\ref{tab:rottemp}, where the
parameters listed are weighted averages).
   }
\label{fig:v27region}
\end{figure}

As an example of the NH$_3$(1,1) spectra and line fits obtained with this procedure we show in 
Figure~\ref{fig:v27region} the spectra obtained toward V27  by integrating the emission
in the three smaller areas, enclosed within dashed contours, shown in Figure~\ref{fig:v11}.
As mentioned already in Section~\ref{sec:trot}, for each sub-region we have used the same 
NH$_3$(2,2) spectrum obtained by integrating over the entire source. In addition, the NH$_3$(1,1)
linewidths have been kept fixed, and equal to the 100m values, during the fits to the 
VLA spectra obtained using 
METHOD NH$_3$(1,1) of CLASS, such as the ones shown Figure~\ref{fig:v27region}.
The total mass of each source can then be computed by adding the masses of the individual sub-regions,
estimated as:

\begin{equation}
M_{\rm cd} = 1.38 \, \frac{\pi}{4} D^2 \, N_{\rm H_2} \, m_{\rm H_2}    
\label{eq:cd}
\end{equation}
where $D$ is the sub-region diameter, 
$m_{\rm H_2}$ is the mass of the H$_2$ molecule,   $N_{\rm H_2}$ is the sub-region averaged column density and
1.38 is the correction factor for the abundance of helium and heavier elements in the interstellar medium.
The source diameter is actually an equivalent diameter calculated as $\sqrt{\Omega_{\rm s}/1.133}$, where 
$\Omega_{\rm s}$ represents the solid angle covered by the individual sub-regions shown in Figures~\ref{fig:v11} 
to \ref{fig:v33}. 
We have evaluated $M_{\rm cd}$ for two possible values of the ammonia abundance, corresponding to
the minimum and maximum values found in Table\,3 of \citet{pillai2006}, 
$X[{\rm NH_3}] = 7 \times 10^{-9}$ and $X[{\rm NH_3}] = 10^{-7}$. The corresponding values of
$M_{\rm cd}$ are listed separately in Table~\ref{tab:phys} (columns 6 and 7).
%
%
%
The particle density has then been calculated as: 
\begin{equation}
n_{\rm H_2} = \frac{3 M_{\rm cd} }
{4 \pi (D/2)^3 \mu \, m_{\rm H}   }
\end{equation}
where $\mu=2.33$ is the mean molecular weight per particle and $m_{\rm H}$ is the mass of 
the hydrogen atom.  The two values of $n_{\rm H_2}$ corresponding to 
the choice of $X[{\rm NH_3}]$ are also listed separately in Table~\ref{tab:phys} (columns 8 and 9).

From the linewidths of the observed transitions and the estimated source
angular diameters  we can also derive the mass required for virial equilibrium.
Assuming the source to be spherical and homogeneous (an assumption which is not, however,
well justified in all of our sources), and neglecting contributions from
magnetic field and surface pressure, the virial mass is given by \citep{maclaren1988}:
%
\begin{equation}
M_{\rm vir}{\rm [M_\odot]} = 0.509 \, d{\rm [kpc]} \, \theta_{\rm s}{\rm [arcsec]} \,
(\Delta V {\rm [km/s]})^2
\label{eq:vir}
\end{equation}
where $d=2.3\,$kpc \citep{chapin2008} is the distance to the four BLAST sources. 
The estimated virial masses (column 5 in Table~\ref{tab:phys}) show that the cores 
are more likely to be gravitionally unstable, even when the largest value of 
$X[{\rm NH_3}] = 10^{-7}$ is used, are V33 and, to a lesser extent, V27.
Source V11 seems to be closer to virial equilibrium, compared to V27 and V33, which
would appear to be consistent with V11 being a proto-stellar core (see Section~\ref{sec:evo}). 
For comparison, the virial masses listed in Table~\ref{tab:phys}
are quite smaller compared to the values estimated by \citet{fon04a}
for comparable sizes. However, these authors did not use their NH$_3$ map and spectra to determine
the virial mass, and they measured much larger linewidths in other molecular tracers.
Therefore, the virial masses of \citet{fon04a} are likely to
reflect the much larger degree of turbulence in HMPOs. 

Furthermore, apart from the various approximations involved
in the calculation of $M_{\rm vir}$, its comparison
with the masses obtained by the fit to the spectral energy distribution 
(with assumed values of the dust mass absorption coefficient at 250\,\micron,
$k_{\rm 250} = 10\,$cm$^2$g$^{-1}$, and a dust emissivity index, $\beta=1.5$,
\citealp{chapin2008})
is further complicated by the fact that the BLAST05 observations were sensitive to cold dust
on a much larger scale ($\ga 1 \,$arcmin). 
%
%
%

The comparison between $M_{\rm tot}$ and $M_{\rm vir}$ in Table~\ref{tab:phys} is 
complicated by the many uncertainty factors. In addition, for a power-law density 
distribution of the type $n_{\rm H_2}(r) \propto r^{-m}$,
the virial mass obtained from Eq.\,(\ref{eq:vir}) must be
multiplied by a factor $\frac{3}{5} \, \frac{5-2m}{3-m}$ (see \citealp{maclaren1988}).
For example, \citet{fon04a} found that $m = 2.3$ in the HMPO IRAS\,23385+6053 and thus
their virial masses had to be multiplied by a factor $\simeq 0.35$. 
%

%

%

\subsection{Comparison with Low-Mass Pre-Stellar Cores}
\label{sec:lms}

To analyze the origin of the V10, V11, V27 and V33 cores, we can compare the properties
of these cores with the well-studied low-mass pre-stellar core L1544 
in the Taurus molecular cloud (at a distance of 140\,pc) that appears to be gravitationally
unstable (for a review see \citealp{doty2005} and references therein). 
The L1544 core has a continuum source with a flux density of 17.4\,Jy at 450\,\micron\ 
\citep{shirley2000}, corresponding to a gas+dust mass of 3.2\,M$_\odot$, and a density of 
$1.5 \times 10^6\,$cm$^{-3}$ \citep{ward1999}.
The N$_2$H$^+$ linewidth toward L1544 is 0.3\,km\,s$^{-1}$, and observations toward 
several low-mass starless cores in NH$_3$ and N$_2$H$^+$ show similar linewidths
for both molecules \citep{tafalla2004}.

In order to compare the physical properties of L1544 with those of the BLAST cores 
listed in Table~\ref{tab:phys}, we note that the peak NH$_3$ abundance determined by 
\citet{tafalla2002} toward L1544 was $4.0 \times 10^{-9}$. Therefore, if we consider the 
physical parameters in Table~\ref{tab:phys} corresponding to the {\it lower} value of
the NH$_3$ abundance ($7.0 \times 10^{-9}$, close enough to that estimated in L1544), 
we see that the number density in the BLAST cores is quite similar to
that of L1544, while their mass is a factor of $\sim 20-100$ higher, and is quite higher
even when one considers  the masses of the smaller sub-regions within 
each source, as described in Section~\ref{sec:mass}. 

As far as the velocity dispersion is concerned, the BLAST cores have NH$_3$ linewidths 
more than double that of L1544, with the exception of V10 which has a linewidth
(0.41\,km\,s$^{-1}$, see Table~\ref{tab:methnh3}) only slightly higher than that
of L1544. This would be consistent with the BLAST cores being more turbulent
and could also possibly hide other systematic internal motions (see the case of V11,
Section~\ref{sec:phys}).

What would a core like L1544 appear in the BLAST05 map of Vulpecula? 
Scaling the 450\,\micron\ L1544 emission to 500\,\micron\ at a distance of 2.3\,kpc, 
we would expect an integrated flux density of about 40\,mJy and thus it would go
undetected in the BLAST05 map of Vulpecula \citep{chapin2008}.  
This is consistent with the four BLAST sources being separated, if they were plotted 
on the $L-M$ diagram  of \citet{molinari2008}, from the low-mass regime, as 
mentioned in Section~\ref{sec:sample}.

Table~\ref{tab:phys} shows that only in the case of V11 the total 
mass inferred from the BLAST observations is significantly larger than the mass
estimate based on NH$_3$ emission, given the NH$_3$ abundance range that we have selected. 
In V27 and V33, the mass range obtained from our observations is consistent with
the BLAST mass estimate.
It is possible that in the case of V11 the ammonia emission results from only a small
fraction of molecular material that is concentrated in a denser and more compact structure.
A similar scenario, for example, has been found by \citet{shepherd2004} toward the
core G192~S3, and it could also be the case in V27 and V33 if the NH$_3$ abundance 
were near the upper limit of the range listed in Table~\ref{tab:phys}. 
In this case, an interesting question would be whether the mass determined from the VLA
NH$_3$ observations represents the {\it total} 
reservoir from which any star formation will 
eventually take place in these cores or, rather, further mass will be accreted from the 
larger mass reservoir detected by BLAST. In the first scenario, it is unlikely that these
cores will produce massive stars, unless the total gas and dust mass is much higher,
as it would result if $X[{\rm NH_3}] << 10^{-7}$. 

%
 \begin{figure}
 \centering
 \includegraphics[width=9.5cm]{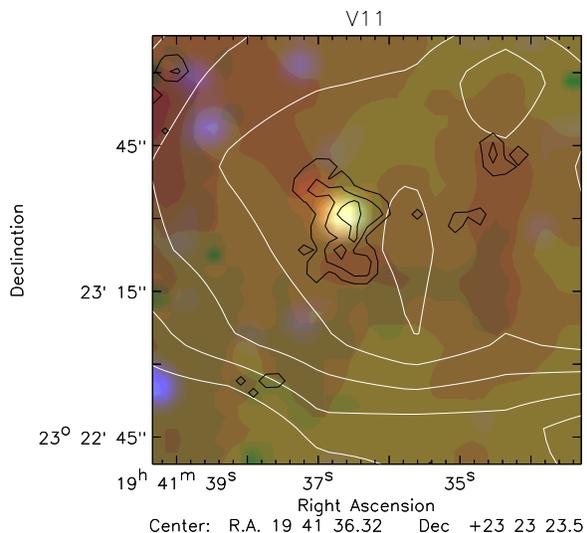}
 \caption{
Three-color {\it Spitzer} image 3.6\,\micron\ (blue), 8\,\micron\ (green), and 24\,\micron\ (red) of
V11 (at the angular resolution of the MIPS 24\,\micron\ map).
The overlaid white, solid contours represent the BLAST 350\,\micron\ emission
(from $4.3\times 10^7$   to  $1.2\times 10^8$ by $1.6\times 10^7$\,Jy\,srad$^{-1}$).
The black, solid contours represent the integrated NH$_3$(1,1) VLA emission (from
0.039   to  0.112 by 0.036\,Jy\,beam$^{-1}$\,km\,s$^{-1}$).
   }
\label{fig:v11color}
\end{figure}

Likewise, one might ask whether the lower-mass fragments in these regions could at a later stage
merge to form a more massive pre-stellar core that could then evolve toward a high-mass star.
Clearly, we cannot currently determine the relative motions of the individual fragments in 
the observed cores. However, we can use the NH$_3$ linewidths as a rough estimate of the 
average turbulent motion, though they do not necessarily represent the relative velocities
of coherent sub-structures within the cores. 
If we take $\sim 0.1$\,pc and $0.8 \sqrt{3}$\,km\,s$^{-1}$ as the
typical fragment (de-projected) separation and 3D velocity dispersion, respectively, we obtain 
a time-scale $\sim 8 \times 10^{4}$\,yr. Although this time-scale is comparable to the accretion 
time-scales estimated by \citet{bonnell2006} for the formation of high-mass stars in the 
competitive accretion scenario, the most significant individual fragments inside 
the cores appear to be more massive than the thermal 
Jeans mass ($\sim 1\,M_\odot$), which is inconsistent with the model of 
\citet{bonnell2006} for competitive accretion. Alternatively, it is quite possible that
we are observing these cores at a phase when merging of low-mass fragments has already
occurred, and that the less significant fragments in Figures~\ref{fig:v10} to \ref{fig:v33}, 
with masses $\ga 1\,M_\odot$, are actually the remnants of this accreting phase.

\subsection{Evolutionary Phase of the Cores}
\label{sec:evo}

As we have seen in Section~\ref{sec:trot} and Section~\ref{sec:mass}, our cores have similar temperatures
but are more compact, and possibly less massive (depending on the NH$_3$
abundance), than IRDCs, which have a typical size $\ga 1\,$pc.
This suggests that they may be similar to the unresolved cores found toward IRDCs,
but in an earlier stage of evolution compared to the sources observed by \citet{mol96}
and \citet{Sridharan05}.

This conclusion is supported by the lack of an {\it IRAS} and {\it MSX} counterpart to these sources,
as mentioned in Section~\ref{sec:sample}. However, we have also inspected the IRAC and MIPS images
and found that only in the case of V11 an IRAC/MIPS point source is
found to be positionally coincident with the NH$_3$(1,1) integrated emission.
As shown in Figure~\ref{fig:v11color} this source has the colors expected for
an embedded protostar. Aperture photometry confirms that the SED of this infrared source is rising
between 3.6 and 24\,\micron. However, we find no point source in the 70\,\micron\ MIPS image.
No other infrared source with similar colors is found within $\simeq 1\,$arcmin from the peak
of either the NH$_3$(1,1) or the BLAST emission. 
Together with the larger linewidth and 
a more systematic velocity shift across the source structure, as discussed in Section~\ref{sec:phys}, 
these properties suggest that V11 is probably the most evolved of all four observed cores, and 
is likely to harbour a protostar. 

In addition, we have also searched for possible radio continuum emission toward the four Vulpecula sources.
In order to do this we have analyzed both the continuum images constructed from the line-free channels of
our VLA observations, and the available maps from the CORNISH survey of the Galactic Plane \citep{purcell2008}.
We do not detect any continuum emission at the level of $\simeq 0.4\,$mJy\,beam$^{-1}$, in our 
line-free channels maps, which is about the same as the noise level achieved at 5\,GHz in the CORNISH maps.
Thus, both the upper limit to the continuum emission (see \citealp{purcell2008}) and the luminosity
estimated from SED fits ($L\sim 50\,L_\odot$, \citealp{chapin2008}) 
imply an ionizing source much less than a B3 zero-age main-sequence star.
The low-luminosity in the radio continuum 
could be because the central object has not yet developed an UC~HII region,
or because of opacity if the surroundings of the protostar (if any) are very dense.

\section{CONCLUSIONS}
\label{sec:concl}

We have observed four candidate high-mass starless cores selected from the BLAST05 survey 
of the Vulpecula region. Our VLA-D observations in the NH$_3$(1,1) and (2,2) lines suggest
that these cores are in very early stages of evolution, prior to the formation of a 
(proto)star or proto-cluster. In only one of these cores, V11, the VLA peak emission of
NH$_3$(1,1) is associated with 
an infrared source having the colors expected for an embedded protostar.
The four cores are cold ($T_{\rm k} < 16\,$K), 
relatively quiescent ($\Delta V \la 0.8 \,$km\,s$^{-1}$) but with a higher velocity
dispersion  compared to e.g., L1544, and show 
a filamentary and clumpy  
structure. Our VLA-D data have limited spectral resolution, but we can clearly observe
a significant velocity substructure within $\sim 1\,$km\,s$^{-1}$.

Based on the comparison with a typical low-mass pre-stellar core, we find that these
BLAST cores are more massive and more luminous. V11 is likely to have already formed an
intermediate-mass (proto)star and given the mass range obtained for V27 and V33 they too
could form intermediate- or even high-mass stars.
However, a confirmation of this conclusion will require
a more accurate determination of the NH$_3$ abundance in these cores. 



\acknowledgments
The authors thank R. Cesaroni and A. Lopez Sepulcre
for kindly providing the 100-m data.
L.O. and C.M.P. acknowledges travel support from NRAO during their stay at the 
Array Operations Center in Socorro (NM).  
PH acknowledges partial support from NSF grant AST-0908901.
PGM acknowledges partial support from NSERC.
The authors also wish to thank an anonymous referee whose comments have 
considerably improved the paper.

\bibliographystyle{apj}
\bibliography{apj-jour,refs}

\end{document}